\documentclass[10pt,journal,compsoc]{IEEEtran}

\usepackage{hyperref}
\usepackage{graphicx}
\usepackage{makecell}
\usepackage{algorithmic}
\usepackage[ruled,vlined]{algorithm2e}
\usepackage{multirow}
\usepackage{amsmath}
\usepackage{pifont}
\usepackage{bm}
\usepackage{amsfonts}
\usepackage{amsthm}
\usepackage{amssymb}
\usepackage{color}
\usepackage{subfigure}
\usepackage[justification=centering]{caption}
\usepackage[font=footnotesize]{caption}

\ifCLASSOPTIONcompsoc
  \usepackage[nocompress]{cite}
\else
  \usepackage{cite}
\fi

\ifCLASSINFOpdf
\else
\fi

\begin{document}

\title{cuFastTuckerPlus: A Stochastic Parallel Sparse FastTucker Decomposition Using GPU Tensor Cores}

\author{	
	Zixuan Li,
	Mingxing Duan,
	Huizhang Luo,
	Wangdong Yang,
	Kenli Li, \emph{Senior Member}, \emph{IEEE},
	Keqin Li, \emph{Fellow}, \emph{IEEE}.
	\IEEEcompsocitemizethanks
	{
		\IEEEcompsocthanksitem
		Zixuan Li, Mingxing Duan, Huizhang Luo, Wangdong Yang and Kenli Li are with the College of Computer Science and Electronic Engineering, Hunan University, and National Supercomputing Center in Changsha, Hunan, Changsha 410082, China.
		\IEEEcompsocthanksitem Keqin Li is with the College of Computer Science and Electronic Engineering,
		Hunan University, and the National Supercomputing Center in Changsha,
		Hunan, Changsha 410082, China, and with the Department of Computer
		Science, State University of New York, New Paltz, NY 12561, USA.
		\IEEEcompsocthanksitem Corresponding author: Kenli Li.
		\protect\\ E-mail:
		zixuanli@hnu.edu.cn,
		duanmingxing@hnu.edu.cn,
		luohuizhang@hnu.edu.cn,
		yangwangdong@hnu.edu.cn,	
		lkl@hnu.edu.cn,
		lik@newpaltz.edu.
	}
}

\markboth{}
{Shell \MakeLowercase{\textit{et al.}}: Bare Advanced Demo of IEEEtran.cls for Journals}
\IEEEtitleabstractindextext{
\begin{abstract}
Sparse tensors are prevalent in real-world applications, often characterized by their large-scale, high-order, and high-dimensional nature. Directly handling raw tensors is impractical due to the significant memory and computational overhead involved. The current mainstream approach involves compressing or decomposing the original tensor. One popular tensor decomposition algorithm is the Tucker decomposition. However, existing state-of-the-art algorithms for large-scale Tucker decomposition typically relax the original optimization problem into multiple convex optimization problems to ensure polynomial convergence. Unfortunately, these algorithms tend to converge slowly. In contrast, tensor decomposition exhibits a simple optimization landscape, making local search algorithms capable of converging to a global (approximate) optimum much faster. In this paper, we propose the FastTuckerPlus algorithm, which decomposes the original optimization problem into two non-convex optimization problems and solves them alternately using the Stochastic Gradient Descent method. Furthermore, we introduce cuFastTuckerPlus, a fine-grained parallel algorithm designed for GPU platforms, leveraging the performance of tensor cores. This algorithm minimizes memory access overhead and computational costs, surpassing the state-of-the-art algorithms. Our experimental results demonstrate that our method achieves a speedup of $3X$ to $5X$ compared to state-of-the-art algorithms.
\renewcommand{\raggedright}{\leftskip=0pt \rightskip=0pt plus 0cm}
\raggedright
\end{abstract}

}

\maketitle
\IEEEdisplaynontitleabstractindextext
\IEEEpeerreviewmaketitle

\ifCLASSOPTIONcompsoc

\section{Introduction}

Multi-order data, represented as tensors, is pervasive in various domains, including social networks \cite{yin2021tensor, wang2021reducing}, recommender systems \cite{wang2019ho, chen2021deep}, cryptography \cite{shin2022passwordtensor}, bioinformatics \cite{huang2021tensor}, and neuroscience \cite{zhu2019measuring, mirzaei2019overlapping}. Tensors serve as higher-order generalizations of matrices, capable of capturing complex relationships among multiple entities or variables. Leveraging tensor decomposition techniques provides a powerful framework for uncovering latent structures, reducing dimensionality, and performing feature extraction \cite{kolda2009tensor}. Through this decomposition process, a concise representation of the data is obtained, enabling efficient storage, analysis, and interpretation. However, in many real-world applications, tensors often exhibit sparsity, where a significant number of elements are zero or missing. Sparsity is a common characteristic in large-scale datasets due to inherent constraints or limitations during data acquisition. For instance, in social network analysis, connections between individuals may be sparse, while in image analysis, pixels representing background regions are predominantly zero. The growing interest in large-scale sparse tensor decomposition in recent years stems from the exponential growth of data in the aforementioned domains. Nonetheless, the analysis of such tensors poses unique challenges due to their immense size, sparsity patterns, and computational complexity.

Sparse tensors, in contrast to dense tensors, exhibit a substantial number of zero or missing entries, which are commonly encountered in real-world applications due to the inherent sparsity of the underlying phenomena. The presence of sparsity in tensors presents unique challenges and opportunities for data analysis and computational algorithms. Traditional tensor decomposition algorithms, originally designed for dense tensors, face difficulties in effectively scaling to large-scale sparse tensors due to the computational overhead involved in processing numerous zero entries. Consequently, specialized algorithms and techniques have been developed to leverage the sparsity patterns and efficiently factorize sparse tensors. Furthermore, sparse tensor decomposition techniques should possess the ability to accurately recover missing entries and handle inherent data noise.

Tucker decomposition is a widely used technique for factorizing tensors into a core tensor and factor matrices, enabling the analysis of complex relationships and latent structures within tensors. There are two primary categories of methods used for implementing Tucker decomposition. The first category is based on Singular Value Decomposition (SVD) methods, such as High Order Singular Value Decomposition (HOSVD) \cite{de2000multilinear} and Higher Order Orthogonal Iteration (HOOI) \cite{de2000best}. The second category is comprised of gradient-based methods, including Alternating Least Squares (ALS) \cite{comon2009tensor}, Coordinate Descent (CD) \cite{xu2013block}, and Stochastic Gradient Descent (SGD) \cite{ge2015escaping}. SVD-based methods typically involve two essential computational steps: Tensor-Time-Matrix-chain (TTMc) and SVD. Both of these steps are computationally expensive and require significant memory usage. Gradient-based methods, on the other hand, calculate the gradient of a parameter and update it using gradient descent or least squares. Sparse Tucker decomposition extends this technique to handle large-scale sparse tensors where a substantial number of elements are zero or missing.

Parallel computing refers to the utilization of multiple processing units or resources that work simultaneously to solve computational problems more efficiently. By parallelizing the Sparse Tucker decomposition process, the computational time can be significantly reduced, enabling the analysis of larger and more complex sparse tensors. These techniques leverage various parallel computing architectures, including multi-core processors, distributed systems, and GPU accelerators, to achieve high-performance factorization of large-scale sparse tensors.
Several parallel algorithms have been developed for Sparse Tucker decomposition. ParTi! \cite{nisa2019load} is an HOOI-based parallel algorithm designed for GPUs, which utilizes the Balanced Compressed Sparse Fiber (B-CSF) format to accelerate TTMC calculations. P-tucker \cite{oh2018scalable} is an ALS-based parallel algorithm for multi-core CPUs, reducing memory requirements for updating factor matrices, but it incurs significant computational overhead and exhibits unbalanced load distribution. Vest \cite{park2021vest} is a CD-based parallel algorithm for multi-core CPUs, which prunes unimportant entries to reduce calculations, but the pruning process itself is time-consuming. SGD\_\_Tucker \cite{li2020sgd} is an SGD-based parallel algorithm for multi-core GPUs, dividing high-dimensional intermediate variables into smaller ones to reduce memory overhead during updates. GTA \cite{oh2019high} is an ALS-based parallel algorithm designed for heterogeneous platforms and serves as an extended version of P-Tucker. cuTucker \cite{li2022cu_fasttucker} is an SGD-based parallel algorithm that runs on multiple GPUs. cuFastTucker \cite{li2022cu_fasttucker} is an SGD-based parallel algorithm for multiple GPUs, decomposing the core tensor into multiple core matrices to reduce space and computational overhead. Building upon cuFastTucker, cuFasterTucker \cite{li2022cufastertucker} further reduces the computation of shared and reusable intermediate variables, maximizing the utilization of GPU computing resources, and introduces non-negative constraints.
Table \ref{algorithms} provides a summary of the above algorithms' ability to handle High-order, High-dimensional, Large-scale, and Sparse Tensors (HHLST), along with their computational and memory overhead, load balancing, and convergence speed. Three levels (high, medium, low) are used to assess an algorithm's adaptability to large-scale, high-order, and high-dimensional tensors, where higher levels indicate higher capabilities. Similarly, three levels (high, medium, low) distinguish the memory and computational complexity of an algorithm, with lower levels indicating lower complexities. The load balancing and convergence speed of algorithms are also evaluated using three levels (high, medium, low), with higher levels implying better load balancing or faster convergence. Among the evaluated algorithms, cuFastTucker, cuFasterTuckerCOO, and cuFasterTucker are capable of handling HHLST.

\begin{table*}[htbp]	
	\caption{The Performance of Parallel Sparse Tensor Decomposition Algorithms.}
	\footnotesize
	\centering
	\label{algorithms}
	\begin{tabular}{c|ccc|cc|ccc}
		\hline \hline
		Algorithm    
		&  \makecell[c]{Large\\Scale}          
		&  \makecell[c]{High\\Order}
		&  \makecell[c]{High\\Dimensional}
		&  \makecell[c]{Computational\\Overhead}
		&  \makecell[c]{Memory\\Overhead}
		&  \makecell[c]{Load\\Balancing} 
		&  \makecell[c]{Convergence\\Speed}
		&  \makecell[c]{Tensor\\Core}  \\
		\hline
		P-Tucker            & Low    & Low 	  & Medium & High	& High 	 & Low	  & Low	   & Low	    \\
		Vest            	& Low    & Low 	  & Medium & High	& High 	 & Low    & Low    & Low	    \\
		SGD\_\_Tucker       & Low  	 & Low 	  & Medium & High	& Medium & High   & Medium & High	\\
		ParTi            	& Medium & Medium & Low	   & High	& High 	 & Medium & Medium & Medium	\\
		GTA            		& Low    & Low 	  & Medium & High	& High 	 & Low	  & Medium & Low	    \\
		cuTucker           	& High   & Low 	  & High   & High	& High 	 & High   & Medium & High	\\
		cuFastTucker        & High   & High   & High   & Medium & Low 	 & High	  & Medium & High	\\
		cuFasterTucker      & High   & High   & High   & Low	& Low 	 & Medium & Medium & Low	    \\
		cuFasterTuckerCOO   & High   & High   & High   & Low	& Low 	 & High	  & Medium & Medium	\\
		\hline 
		cuFastTuckerPlus    & High   & High   & High   & Low	& Low    & High   & High   & High	\\
		\hline \hline
	\end{tabular}
\end{table*}

The aforementioned algorithms are based on convex relaxations of non-convex optimization problems, transforming them into convex optimization problems. While they can achieve convergence in polynomial time, they often exhibit slow convergence in practice. In contrast, local search algorithms have demonstrated fast convergence in practical scenarios \cite{ma2021local}. Many non-convex optimization problems are conjectured to possess favorable geometric properties, wherein all local optima are (approximately) global optima \cite{dauphin2014identifying, choromanska2015loss}. Tensor factorization \cite{ge2015escaping}, matrix awareness \cite{park2017non, bhojanapalli2016global}, and matrix completion \cite{ge2016matrix} exhibit good optimization landscapes, where all locally optimal solutions are globally optimal. Similarly, common low-rank matrix factorizations share a unified optimization landscape, wherein all local optima are global optima, and high-order saddle points are absent \cite{ge2017no}. Moreover, for an exact standard Tucker decomposition, all locally optimal solutions are also globally optimal \cite{frandsen2022optimization}. These problems can be effectively addressed using fundamental optimization algorithms such as stochastic gradient descent \cite{ge2015escaping, carmon2018accelerated, agarwal2016finding, jin2017escape}. From these observations, we infer that FastTucker decomposition also possesses a favorable optimization landscape, where all local optima are (approximately) global optima, and can be effectively solved using stochastic gradient descent.

Tensor Cores have emerged as specialized hardware components that provide powerful acceleration for tensor computations. They leverage the parallelism and computational capabilities of modern GPUs to efficiently perform tensor operations. Tensor Cores incorporate dedicated hardware units and optimized algorithms, enabling rapid and precise tensor computations. The introduction of Tensor Cores has revolutionized the field of tensor computations, delivering significant speedups in various applications, including deep neural network training \cite{feng2021apnn} and complex physical system simulations \cite{finkelstein2021quantum}. The efficient handling of large-scale tensor computations by Tensor Cores has opened up new possibilities for addressing challenging problems in machine learning \cite{huang2022high, li2020accelerating, zhu2019sparse}, scientific simulations \cite{zachariadis2020accelerating, mukunoki2020dgemm}, and data analysis \cite{firoz2020feasibility}. However, not all algorithms are suitable for acceleration using Tensor Cores. Table \ref{algorithms} summarizes the adaptability of above algorithms to Tensor Core acceleration, categorized into three levels: High, Medium, and Low. A higher level indicates a greater suitability for Tensor Core acceleration.

In this paper, we introduce cuFastTuckerPlus, a parallel sparse FastTucker decomposition algorithm designed for the GPU platform with Tensor Cores. Unlike previous convex optimization approaches, cuFastTuckerPlus utilizes a non-convex stochastic optimization strategy. The algorithm consists of two parts: simultaneous updates of all factor matrices followed by simultaneous updates of all core matrices. As depicted in Table \ref{algorithms}, cuFastTuckerPlus demonstrates strong performance across various metrics. Our contributions can be summarized as follows:

\begin{enumerate}
	
	\item \emph{Algorithm.} 
	The proposed method, cuFastTuckerPlus, is a stochastic optimization strategy specifically developed for parallel sparse FastTucker decomposition on the GPU platform using Tensor Cores. It addresses the optimization problem by decomposing it into two non-convex optimization subproblems, which are alternately solved to achieve convergence. The algorithm exhibits rapid convergence and effectively utilizes the capabilities of Tensor Cores, resulting in faster performance compared to existing state-of-the-art (SOTA) parallel Tucker decomposition algorithms such as cuFastTucker and cuFasterTucker. Although cuFastTuckerTC and cuFasterTuckerTC are versions of cuFastTucker and cuFasterTucker respectively, accelerated by Tensor Cores, their performance does not match that of cuFastTuckerPlus.
	
	\item \emph{Theory.}	
	In comparison to cuFastTucker, our proposed cuFastTuckerPlus exhibits a smaller memory access overhead of $(M+R)\sum_{n=1}^{N}J_n$ and a computational overhead of $MR\big(\sum_{n=1}^{N}J_n+N(N-2)\big)$ for key steps. On the other hand, cuFasterTucker aims to reduce the real-time computation of $\textbf{C}_{\Psi^{(n)},:}^{(n)}$, $n$ $\in$ $\{N\}$, which incurs an overhead of $MR\sum{n=1}^{N}J_n$, by introducing an additional memory access of $\textbf{C}_{\Psi^{(n)},:}^{(n)}$, $n$ $\in$ $\{N\}$, resulting in an overhead of $N(N-1)R$. In the case of cuFastTuckerPlus, the calculation of $\textbf{C}_{\Psi^{(n)},:}^{(n)}$, $n$ $\in$ $\{N\}$ is performed in real-time using Tensor Cores, without introducing any additional memory access overhead. Furthermore, the increased running time of cuFastTuckerPlus is lower than the additional memory access time required for $\textbf{C}_{\Psi^{(n)},:}^{(n)}$, $n$ $\in$ $\{N\}$. These advantages make cuFastTuckerPlus superior to cuFasterTucker in terms of both computation and memory access.
	
	\item \emph{Performance.} 
	The experimental results of cuFastTuckerPlus demonstrate its superior convergence speed and efficiency compared to the current SOTA algorithms. Specifically, during the factor matrix update step, cuFastTuckerPlus achieves a speedup of $18X$ to $22X$ compared to cuFastTucker and $2X$ to $3X$ compared to cuFasterTucker. Similarly, during the core matrix update step, cuFastTuckerPlus achieves a speedup of $43X$ to $44X$ compared to cuFastTucker and $4X$ to $5X$ compared to cuFasterTucker. These significant speed improvements indicate the effectiveness and efficiency of cuFastTuckerPlus in terms of both computational time and memory access.
	
\end{enumerate}

The code for cuFastTuckerPlus, as utilized in this paper, along with a toy dataset, is available for reproducibility at \href{https://github.com/ZixuanLi-China/cuFastTuckerPlus}{here}. The subsequent sections of this paper are structured as follows. Section \ref{Section_Preliminaries} introduces the notations, definitions, and the problem to be addressed, along with its basic algorithm. Section \ref{Section_Proposed_Method} presents our proposed method, which is a non-convex optimization algorithm called FastTuckerPlus decomposition. Section \ref{Section_cuFastTuckerPlus_On_GPU} describes our proposed fine-grained parallel sparse Tucker algorithm, cuFasterTuckerPlus, designed specifically for the GPU platform with Tensor Cores. Section \ref{Section_Experiments} showcases the experimental results of cuFastTuckerPlus, comparing its performance to SOTA algorithms. Finally, in Section \ref{Section_Conclusion}, we summarize our work.

\section{Preliminaries}\label{Section_Preliminaries}

In this paper, Section \ref{Section_Notations} provides a comprehensive description of the notations used. Section \ref{Section_Basic_Definitions} presents the fundamental definitions necessary for understanding the concepts discussed. The specific problem to be addressed is outlined in Section \ref{Section_Problem}, while Section \ref{Section_SGD-based_Sparse_FastTucker_Decomposition_Algorithm} elaborates on the SGD-based convex optimization method proposed to tackle this problem. For easy reference, Table \ref{Table Notations} summarizes the key notations utilized throughout the paper.

\begin{table}[!htbp]
	\setlength{\abovedisplayskip}{0pt}
	\setlength{\belowdisplayskip}{0pt}
	\renewcommand{\arraystretch}{1.5}
	\caption{Table of symbols.}
	\centering
	\footnotesize
	\label{Table Notations}
	\tabcolsep1pt
	\begin{tabular}{cc}
		\hline
		\makecell[c]{Symbol}                         & \makecell[c]{Definition}\\
		\hline
		\makecell[c]{$\bm{\mathcal{X}}$}             & \makecell[c]{The input $N$-order tensor $\in$ $\mathbb{R}^{I_{1}\times I_{2}\times\cdots \times I_{N}}$}\\
		\makecell[c]{$\bm{\mathcal{G}}$}             & \makecell[c]{The $N$-order core tensor $\in$ $\mathbb{R}^{J_{1}\times J_{2}\times\cdots \times J_{N}}$}\\
		\makecell[c]{$x_{i_{1},i_{2},\cdots,i_{n}}$} & \makecell[c]{$i_{1},i_{2},\cdots,i_{n}$th element of tensor $\mathcal{X}$}\\
		\makecell[c]{$\{N\}$}                        & \makecell[c]{The index set $\{1,2,\cdots,N-1,N\}$}\\
		\makecell[c]{$\Omega$}                       & \makecell[c]{The set of non-zero elements in $\bm{\mathcal{X}}$}\\	
		\makecell[c]{$|\Omega|$}                     & \makecell[c]{The number of non-zero elements in $\Omega$}\\	
		\makecell[c]{$\Psi$}  						 & \makecell[c]{The sample set from $\Omega$}\\
		\makecell[c]{$\textbf{A}^{(n)}$}             & \makecell[c]{The $n$th factor matrix $\in$ $\mathbb{R}^{I_{n}\times J_{n}}$}\\
		\makecell[c]{$\textbf{B}^{(n)}$}             & \makecell[c]{The $n$th core matrix $\in$ $\mathbb{R}^{J_{n}\times R}$}\\		
		\makecell[c]{$\textbf{a}_{i_{n}, :}^{(n)}$}  & \makecell[c]{The $i_{n}$th row vector $\in$ $\mathbb{R}^{J_{n}}$ of $\textbf{A}^{(n)}$}\\
		\makecell[c]{$\textbf{b}_{:,r}^{(n)}$}       & \makecell[c]{The $r$th column vector $\in$ $\mathbb{R}^{J_{n}}$ of $\textbf{B}^{(n)}$}\\		
		\makecell[c]{$\circ$}                        & \makecell[c]{Outer product}\\		
		\makecell[c]{$\times_{(n)}$}                 & \makecell[c]{$n$-Mode Tensor-Matrix product}\\
		\makecell[c]{$\odot$}                        & \makecell[c]{$R$ Dot Product}\\
		\makecell[c]{$*$}                      		 & \makecell[c]{Hadamard Product}\\
		\makecell[c]{$\circledast$}                  & \makecell[c]{$R$ Hadamard Product}\\	
		\hline
	\end{tabular}
\end{table}

\subsection{Notations}\label{Section_Notations}

The notation conventions used in this paper are as follows: tensors are represented using bold Euler script letters (e.g., $\bm{\mathcal{X}}$), matrices are denoted by bold uppercase letters (e.g., $\textbf{A}$), vectors are represented using bold lowercase letters (e.g., $\textbf{a}$), and scalars are denoted by regular lowercase or uppercase letters (e.g., $k$ and $N$). The elements of a tensor are specified by combining the symbolic name of the tensor with the corresponding indices. For instance, $x_{i_{1},\cdots,i_{n}}$ denotes the element located at indices $(i_{1},\cdots,i_{n})$ in the tensor $\bm{\mathcal{X}}$. Additionally, $\textbf{a}_{i,:}$ refers to the $i$-th row of matrix $\textbf{A}$, and $\textbf{b}_{:,r}$ denotes the $r$-th column of matrix $\textbf{B}$.

\subsection{Basic Definitions}\label{Section_Basic_Definitions}
\newtheorem{definition}{Definition}
\begin{definition}[$n$-Mode Tensor-Matrix product]
	Given a $N$-order tensor $\bm{\mathcal{G}}$ $\in$ $\mathbb{R}^{J_{1}\times\cdots\times J_{N}}$ and a matrix $\textbf{A}$ $\in$ $\mathbb{R}^{I_{n}\times J_{n}}$,
	$n$-Mode Tensor-Matrix product projects $\bm{\mathcal{G}}$ and $\textbf{A}$ to a new tensor $(\bm{\mathcal{G}}\times_{(n)} \textbf{A})$ $\in$ $\mathbb{R}^{I_{1}\times\cdots \times J_{n-1}\times I_{n}\times J_{n+1} \times\cdots  J_{N}}$ according to the coordinates, where $(\bm{\mathcal{G}}\times_{(n)} \textbf{A})_{j_{1}\times\cdots \times j_{n-1}\times i_{n}\times j_{n+1}\times\cdots\times  j_{N}}$ $=$ $\sum\limits_{j_{n}=1}^{J_{n}}$ $g_{j_{1},\cdots, j_{n},\cdots,j_{N}}$ $\cdot a_{i_{n},j_{n}}$.
\end{definition}

\begin{definition}[$R$ Kruskal Product]
	Given $N$ matrices $\textbf{B}^{(n)}$ $\in$ $\mathbb{R}^{J_{n}\times R}$, $n \in \{N\}$,
	$R$ Kruskal Product projects the $N$ matrices $\textbf{B}^{(n)}$ to a new $N$-order tensor $\widehat{\bm{\mathcal{G}}}$ $\in$ $\mathbb{R}^{J_{1}\times\cdots \times J_{N}}$ where $\widehat{\bm{\mathcal{G}}}$$=\sum_{r=1}^{R} \textbf{b}^{(1)}_{:,r}$$\circ\cdots\circ \textbf{b}^{(n)}_{:,r}\circ\cdots$ $\circ$ $\textbf{b}^{(N)}_{:,r}$.
\end{definition}

\begin{definition}[$R$ Dot Product]
	Given a matrix $\textbf{A}$ $\in$ $\mathbb{R}^{M\times R}$,
	and a matrix $\textbf{B}$ $\in$ $\mathbb{R}^{R\times M}$,
	$R$ Dot Product projects $\textbf{A}$ and $\textbf{B}$ to a new matrice $(\textbf{A} \odot \textbf{B})$
	$\in$ $\mathbb{R}^{M\times 1}$ according to the coordinates, where $(\textbf{A} \odot \textbf{B})_{m,1}$ $=$ $\textbf{a}_{m,:}$ $\cdot$ $\textbf{b}_{:,m}$.	
\end{definition}

\begin{definition}[Hadamard Product]
	Given a matrix $\textbf{A}$ $\in$ $\mathbb{R}^{M \times N}$ and a matrix $\textbf{B}$ $\in$ $\mathbb{R}^{M \times N}$, 
	Hadamard Product projects $\textbf{A}$ and $\textbf{B}$ to a new matrice $(\textbf{A} * \textbf{B})$
	$\in$ $\mathbb{R}^{M\times N}$ according to the coordinates, where $(\textbf{A} * \textbf{B})_{m,n}$ $=$ $a_{m,n}$ $\cdot$ $b_{m,n}$.
\end{definition}

\begin{definition}[$R$ Hadamard Product]
	Given a matrix $\textbf{A}$ $\in$ $\mathbb{R}^{M \times 1}$ and a matrix $\textbf{B}$ $\in$ $\mathbb{R}^{M \times R}$, 
	Hadamard Product projects $\textbf{A}$ and $\textbf{B}$ to a new matrice $(\textbf{A} \circledast \textbf{B})$
	$\in$ $\mathbb{R}^{M\times R}$ according to the coordinates, where $(\textbf{A} \circledast \textbf{B})_{:,n}$ $=$ $\textbf{A}$ $*$ $\textbf{B}_{:,r}$.
\end{definition}

\subsection{Problem}\label{Section_Problem}

Consider a sparse and incomplete $N$-order tensor $\bm{\mathcal{X}} \in \mathbb{R}^{I_{1}\times\cdots \times I_{N}}$. Let $\Omega$ denote the set of non-zero elements in $\bm{\mathcal{X}}$, and $|\Omega|$ represent the number of non-zero elements in $\Omega$. The goal of Tensor Completion is to estimate the missing elements in $\bm{\mathcal{X}}$ based on the available non-zero elements in $\Omega$. Sparse Tucker Decomposition is a widely used method for tensor completion, which involves finding $N$ factor matrices $\textbf{A}^{(n)}$ $\in$ $\mathbb{R}^{I_{n}\times J_{n}}$, $n \in \{N\}$ along with an $N$-order core tensor $\bm{\mathcal{G}}$ $\in$ $\mathbb{R}^{J_{1}\times\cdots\times J_{N}}$. The aim is to approximate the elements $x_{i_1,\dots,i_N}$ of the tensor $\bm{\mathcal{X}}$ by $\widehat{x}_{i_1,\dots,i_N}$ given by 
\begin{equation}\label{tucker}
	\tiny
	\begin{aligned}
		x_{i_1,\dots,i_N}\approx\widehat{x}_{i_1,\dots,i_N}=\bm{\mathcal{G}}\times_{(1)}\textbf{a}^{(1)}_{i_1,:}\times_{(2)}\cdots\times_{(n)}\textbf{a}^{(n)}_{i_n,:}\times_{(n+1)}\cdots\times_{(N)}\textbf{a}^{(N)}_{i_N,:}
	\end{aligned}
\end{equation}                               
and the missing elements can be predicted using Equation (\ref{tucker}).

Sparse FastTucker Decomposition is a variation of the Tucker Decomposition. 
It uses $N$ core matrices $\textbf{B}^{(n)}$ $\in$ $\mathbb{R}^{J_{n}\times R}$, $n \in \{N\}$ to approximate the core tensor $\bm{\mathcal{G}}$ by: 
\begin{equation}\label{fasttucker_core_tensor}
	\tiny
	\begin{aligned}
		\bm{\mathcal{G}}\approx\widehat{\bm{\mathcal{G}}}=\sum_{r=1}^{R} \textbf{b}^{(1)}_{:,r}\circ\cdots\circ\textbf{b}^{(N)}_{:,r}
	\end{aligned}
\end{equation}
which can reduce the space and computational complexity.
In sum, the Sparse FastTucker decomposition is to find $N$ factor matrices $\textbf{A}^{(n)}$ $\in$ $\mathbb{R}^{I_{n}\times J_{n}}$, $n \in \{N\}$ and $N$ core matrices $\textbf{B}^{(n)}$ $\in$ $\mathbb{R}^{J_{n}\times R}$, $n \in \{N\}$, such that
\begin{equation}\label{fasttucker}
	\tiny
	\begin{aligned}
		x_{i_1,\dots,i_N}\approx\widehat{x}_{i_1,\dots,i_N}=&\bigg(\sum_{r=1}^{R} \textbf{b}^{(1)}_{:,r}\circ\cdots\circ \textbf{b}^{(n)}_{:,r}\circ\cdots\circ \textbf{b}^{(N)}_{:,r}\bigg)\times_{(1)}\textbf{a}^{(1)}_{i_1,:}\times_{(2)}\cdots\\
		&\times_{(n)}\textbf{a}^{(n)}_{i_n,:}\times_{(n+1)}\cdots\times_{(N)}\textbf{a}^{(N)}_{i_N,:}\\
		=&\sum_{r=1}^{R} (\textbf{a}^{(1)}_{i_1,:}\textbf{b}^{(1)}_{:,r})\cdots(\textbf{a}^{(N)}_{i_N,:}\textbf{b}^{(N)}_{:,r})\\
		=&\sum_{r=1}^{R}\prod_{n=1}^{N}c^{(n)}_{i_n,r}
	\end{aligned}
\end{equation}
where $\textbf{C}^{(n)}$ $=$ $\textbf{A}^{(n)}\textbf{B}^{(n)}$ $\in$ $\mathbb{R}^{I_{n}\times R}$, and $c^{(n)}_{i_n,r}$
is the $(i_n,r)$th element of the $\textbf{C}^{(n)}$.
That is to solve the following non-convex optimization problem:
\begin{equation}\label{low_rank_optimization_all}
	\tiny
	\begin{aligned}
		\mathop{\arg\min}_{\{\textbf{A}^{(n)}\},\{\textbf{B}^{(n)}\}}
		f\bigg(\bm{\mathcal{X}}, \big\{\textbf{A}^{(n)}\big\}, \big\{\textbf{B}^{(n)}\big\} \bigg)
		=&\sum_{x_{i_1,\dots,i_N}\in\bm{\mathcal{X}}}\bigg\|x_{i_1,\dots,i_N}- \widehat{x}_{i_1,\dots,i_N}\bigg\|_{2}^{2}\\
		&+\lambda_{\textbf{A}}\|\textbf{A}^{(n)}\|_{2}^{2}+\lambda_{\textbf{B}}\|\textbf{B}^{(n)}\|_{2}^{2}
	\end{aligned}
\end{equation}
which adds regularization terms to prevent over fitting. 
Both $\lambda_{\textbf{A}}$ and $\lambda_{\textbf{B}}$ are hyperparameters.

\subsection{SGD-based Sparse FastTucker Decomposition Algorithm}
\label{Section_SGD-based_Sparse_FastTucker_Decomposition_Algorithm}

SGD is a commonly employed method for large-scale optimization problems. It enables the partitioning of a large task into smaller subtasks, making it highly amenable to parallelization. Additionally, SGD mitigates the issue of excessive memory consumption by avoiding the need to process the entire dataset at once.

In \ref{SGD}, a sample set $\Psi$, consisting of $M$ randomly selected samples from the total sample set $\Omega$, is utilized. This selection ensures that the computational cost and memory requirements remain manageable. By using a subset of the data, SGD provides an approximation to the true gradient and facilitates efficient optimization.
\begin{equation}
	\tiny
	\begin{aligned}\label{SGD}
		w&\leftarrow w-\gamma\frac{\partial f_{\Psi}(w)}{\partial w}=w-\gamma\frac{1}{M}\sum_{i\in\Psi}\frac{\partial f_{i}(w)}{\partial w}
	\end{aligned}
\end{equation}

On the other hand, the non-convex global optimization objective (\ref{low_rank_optimization_all}) can be transformed into $N$ convex single-sample factor matrix optimization objectives:
\begin{equation}\label{low_rank_optimization_single}
	\tiny
	\begin{aligned}
		\mathop{\arg\min}_{\textbf{a}^{(n)}_{i_{n},:}}
		f\bigg(\textbf{a}^{(n)}_{i_{n},:}\bigg|x_{i_1,\dots,i_N}, \big\{\textbf{A}^{(n)}\big\}, \big\{\textbf{B}^{(n)}\big\} \bigg)
		=\bigg\|x_{i_1,\dots,i_N}- \widehat{x}_{i_1,\dots,i_N}\bigg\|_{2}^{2}
		+\lambda_{\textbf{A}}\|\textbf{a}^{(n)}_{i_{n},:}\|_{2}^{2}
	\end{aligned}
\end{equation}
for all $i_n \in\{I_n\}$, $n \in\{N\}$, and $N$ convex single-sample core matrix optimization objectives:
\begin{equation}\label{core_tensor_optimization_single}
	\tiny
	\begin{aligned}
		\mathop{\arg\min}_{\textbf{B}^{(n)}}
		f\bigg(\textbf{B}^{(n)}\bigg|x_{i_1,\dots,i_N}, \big\{\textbf{A}^{(n)}\big\}, \big\{\textbf{B}^{(n)}\big\} \bigg)
		=\bigg\|x_{i_1,\dots,i_N}- \widehat{x}_{i_1,\dots,i_N}\bigg\|_{2}^{2}
		+\lambda_{\textbf{B}}\|\textbf{B}^{(n)}\|_{2}^{2}
	\end{aligned}
\end{equation}
for all $n \in\{N\}$. 
For any $n \in \{N\}$, with fixed $\textbf{A}^{(k)}$, $k\neq n$, $k \in \{N\}$ and $\textbf{B}^{(k)}$, $k \in \{N\}$, the optimization objective (\ref{low_rank_optimization_single}) is convex.
Similarly, for any $n \in \{N\}$, with fixed $\textbf{A}^{(k)}$, $k \in \{N\}$ and $\textbf{B}^{(k)}$, $k\neq n$, $k \in \{N\}$, the optimization objective (\ref{core_tensor_optimization_single}) is convex.

The SGD for the optimized function $f\bigg(\textbf{a}^{(n)}_{i_{n},:}$ $\bigg|x_{i_1,\dots,i_N},$ $\big\{\textbf{A}^{(n)}\big\},$ $\big\{\textbf{B}^{(n)}\big\}, n \in\{N\} \bigg)$ is derived as follows:
\begin{equation}\label{Gradient_low_rank_cp}
	\tiny
	\begin{aligned}
		&\frac{\partial f\bigg(\textbf{a}^{(n)}_{i_{n},:}\bigg|x_{i_1,\dots,i_N}, \big\{\textbf{A}^{(n)}\big\},\big\{\textbf{B}^{(n)}\big\}, n \in\{N\} \bigg)}{\partial \textbf{a}^{(n)}_{i_{n},:}}\\
		=&-\bigg(x_{i_1,\dots,i_N}-\sum_{r=1}^{R}\prod_{n=1}^{N}(\textbf{a}^{(n)}_{i_n,:}\textbf{b}^{(n)}_{:,r})\bigg)\sum_{r=1}^{R}\big(\textbf{b}^{(n)}_{:,r}\cdot\prod_{k\neq n}^{N}(\textbf{a}^{(k)}_{i_{k,:}}\textbf{b}^{(k)}_{:,r})\big)^{T}
		+\lambda_{\textbf{A}}\textbf{a}^{(n)}_{i_{n},:}\\
		=&-\bigg(x_{i_1,\dots,i_N}-\textbf{a}^{(n)}_{i_{n},:}\textbf{B}^{(n)}\textbf{d}^{(n)^{T}}_{i_{n},:}\bigg)\textbf{d}^{(n)}_{i_{n},:}\textbf{B}^{(n)^{T}}+\lambda_{\textbf{A}}\textbf{a}^{(n)}_{i_{n},:}\\
	\end{aligned}
\end{equation}
and the SGD for the optimized function $f\bigg(\textbf{B}^{(n)}\bigg|x_{i_1,\dots,i_N},$ $\big\{\textbf{A}^{(n)}\big\},$$\big\{\textbf{B}^{(n)}\big\},n \in\{N\}\bigg)$ is derived as follows:
\begin{equation}\label{Gradient_core_tensor_cp}
	\tiny
	\begin{aligned}
		&\frac{\partial f\bigg(\textbf{B}^{(n)}\bigg|x_{i_1,\dots,i_N}, \big\{\textbf{A}^{(n)}\big\},\big\{\textbf{B}^{(n)}\big\},n \in\{N\}\bigg)}{\partial \textbf{B}^{(n)}}\\
		=&-\bigg(x_{i_1,\dots,i_N}-\sum_{r=1}^{R}\prod_{k=1}^{N}(\textbf{a}^{(k)}_{i_k,:}\textbf{b}^{(k)}_{:,r}\big)\bigg)\textbf{a}^{(n)^{T}}_{i_{n},:}\bigg((\textbf{a}^{(1)}_{i_{1},:}\textbf{B}^{(1)})\\
		&*\cdots*(\textbf{a}^{(n-1)}_{i_{n-1},:}\textbf{B}^{(n-1)})*(\textbf{a}^{(n+1)}_{i_{n+1},:}\textbf{B}^{(n+1)})*\cdots*(\textbf{a}^{(N)}_{i_{N},:}\textbf{B}^{(N)})\bigg)+\lambda_{\textbf{B}}\textbf{B}^{(n)}\\
		=&-\bigg(x_{i_1,\dots,i_N}-\textbf{c}^{(n)}_{i_{n},:}\textbf{d}^{(n)^{T}}_{i_{n},:}\bigg)\textbf{a}^{(n)^{T}}_{i_{n},:}\textbf{d}^{(n)}_{i_{n},:}+\lambda_{\textbf{B}}\textbf{B}^{(n)}\\
	\end{aligned}
\end{equation}

In our approach, we define $\textbf{d}^{(n)}_{i_{n},:}$ as the notation for the product of $\textbf{c}_{i_1,:}^{(1)}$ $*$ $\cdots$ $*$ $\textbf{c}_{i_{n-1},:}^{(n-1)}$ $*$ $\textbf{c}_{i_{n+1},:}^{(n+1)}$ $*$ $\cdots$ $*$ $\textbf{c}_{i_N,:}^{(N)}$ for the element $x_{i_1,\dots,i_N}$.

\begin{algorithm}[hptb]
	\tiny
	\caption{Sparse FastTucker Decomposition}
	\label{Algorithm_FastTucker}
	\vspace{.1cm}
	$\textbf{Input}$: Sparse tensor $\mathcal{X}$ $\in$ $\mathbb{R}^{I_{1}\times\cdots\times I_{N}}$, 
	ranks $J_{n}$, $n \in \{N\}$ and $R$,
	learning rates $\gamma_{\textbf{A}}$ and $\gamma_{\textbf{B}}$,
	regularization parameters $\lambda_{\textbf{A}}$ and $\lambda_{\textbf{B}}$, and iterations $T$.\\
	$\textbf{Output}$: Factor matrices $\textbf{A}^{(n)}$, $n$ $\in$ $\{N\}$ and core matrices $\textbf{B}^{(n)}$, $n$ $\in$ $\{N\}$.\\
	\begin{algorithmic}[1]
		\STATE Initialize factor matrices $\textbf{A}^{(n)}$ $\in$ $\mathbb{R}^{I_{n}\times J_{n}}$, $n \in \{N\}$ and core 
		matrices $\textbf{B}^{(n)}$ $\in$ $\mathbb{R}^{J_{n}\times R}$, $n \in \{N\}$.
		\FOR{$t$ from $1$ to $T$}
		\FOR{$n$ from $1$ to $N$}
		\FOR{$i_n$ from $1$ to $I_n$}
		\FOR{randomly takes $\Psi$ from $\Omega_{i_{n}}^{(n)}$}
		\STATE Update $\textbf{a}^{(n)}_{i_{n},:}$ by equations (\ref{SGD}) and (\ref{Gradient_low_rank_cp}).
		\ENDFOR
		\ENDFOR
		\ENDFOR
		\FOR{$n$ from $1$ to $N$}
		\FOR{randomly takes $\Psi$ from $\Omega$}
		\STATE Update $\textbf{B}^{(n)}$ by equations (\ref{SGD}) and (\ref{Gradient_core_tensor_cp}).
		\ENDFOR
		\ENDFOR	
		\ENDFOR		
	\end{algorithmic}
\end{algorithm}

Algorithm \ref{Algorithm_FastTucker} presents the basic SGD-based Sparse FastTucker algorithm, which achieves convergence by sequentially updating $\textbf{A}^{(n)}$, $n \in \{N\}$ and $\textbf{B}^{(n)}$, $n \in \{N\}$. Each update of $\textbf{A}^{(n)}$ or $\textbf{B}^{(n)}$ corresponds to a convex optimization problem, ensuring the algorithm's overall convergence. It is important to note that when updating $\textbf{a}^{(n)}_{i_{n},:}$, the $n$-th index of all non-zero elements in $\Psi$ should be $i_{n}$. We denote the subset of non-zero elements in $\Omega$ whose $n$-th index is $i_{n}$ as $\Omega_{i_{n}}^{(n)}$, which is used for selecting $\Psi$ to match the update of $\textbf{a}^{(n)}_{i_{n},:}$.
Algorithm \ref{Algorithm_FasterTucker} describes the Sparse FasterTucker Decomposition algorithm, which is an improved version of Algorithm \ref{Algorithm_FastTucker}. It introduces intermediate matrices  $\textbf{C}^{(n)}$, $n$ $\in$ $\{N\}$ to store and directly read $\textbf{c}_{i_n,:}^{(n)}$, $i_n$ $\in$ $\{I_n\}$, $n$ $\in$ $\{N\}$ from memory instead of recomputing them. Moreover, it selects $\Psi$ from $\Omega_{i_1,\cdots,i_{n-1},i_{n+1},\cdots,i_N}^{(n)}$. Here, $\Omega_{i_1,\cdots,i_{n-1},i_{n+1},\cdots,i_N}^{(n)}$ represents the subset of non-zero elements in $\Omega$ whose indices in orders $(1, \cdots, n-1, n+1, \cdots, N)$ are $(i_{1}, \cdots, i_{n-1}, i_{n+1}, \cdots, i_{N})$. Thus, for any $x_{i_1,\dots,i_N}$ $\in$ $\Psi$, its corresponding $\textbf{d}^{(n)}_{i_{n},:}$ remains the same and needs to be computed only once.

\begin{algorithm}[hptb]
	\tiny
	\caption{Sparse FasterTucker Decomposition}
	\label{Algorithm_FasterTucker}
	\vspace{.1cm}
	$\textbf{Input}$: Sparse tensor $\mathcal{X}$ $\in$ $\mathbb{R}^{I_{1}\times\cdots\times I_{N}}$, 
	ranks $J_{n}$, $n \in \{N\}$ and $R$,
	learning rates $\gamma_{\textbf{A}}$ and $\gamma_{\textbf{B}}$,
	regularization parameters $\lambda_{\textbf{A}}$ and $\lambda_{\textbf{B}}$, and iterations $T$.\\
	$\textbf{Output}$: Factor matrices $\textbf{A}^{(n)}$, $n$ $\in$ $\{N\}$ and core matrices $\textbf{B}^{(n)}$, $n$ $\in$ $\{N\}$.\\	
	\begin{algorithmic}[1]
		\STATE Initialize factor matrices $\textbf{A}^{(n)}$ $\in$ $\mathbb{R}^{I_{n}\times J_{n}}$, $n \in \{N\}$ and core matrices $\textbf{B}^{(n)}$ $\in$ $\mathbb{R}^{J_{n}\times R}$, $n \in \{N\}$.
		\STATE Calculate and store $\textbf{C}^{(n)},$ $n \in \{N\}$.
		\FOR{$t$ from $1$ to $T$}
		\FOR{$n$ from $1$ to $N$}
		\FOR{$i_1,\cdots,i_{n-1},i_{n+1},\cdots,i_N$ from $1$ to $I_1,\cdots,I_{n-1},I_{n+1},\cdots,I_N$}
		\FOR{randomly takes $\Psi$ from $\Omega_{i_1,\cdots,i_{n-1},i_{n+1},\cdots,i_N}^{(n)}$}
		\FOR{$x_{i_1,\dots,i_{N}}$ $\in$ $\Psi$}
		\STATE Update $\textbf{a}^{(n)}_{i_{n},:}$ by equations (\ref{SGD}) and (\ref{Gradient_low_rank_cp}).
		\ENDFOR
		\ENDFOR
		\ENDFOR
		\STATE Update $\textbf{C}^{(n)}$.
		\ENDFOR
		\FOR{$n$ from $1$ to $N$}
		\FOR{$i_1,\cdots,i_{n-1},i_{n+1},\cdots,i_N$ from $1$ to $I_1,\cdots,I_{n-1},I_{n+1},\cdots,I_N$}
		\FOR{randomly takes $\Psi$ from $\Omega_{i_1,\cdots,i_{n-1},i_{n+1},\cdots,i_N}^{(n)}$}
		\STATE Update $\textbf{B}^{(n)}$ by equations (\ref{SGD}) and (\ref{Gradient_core_tensor_cp}).
		\ENDFOR
		\ENDFOR	
		\STATE Update $\textbf{C}^{(n)}$.
		\ENDFOR	
		\ENDFOR	
	\end{algorithmic}
\end{algorithm}

\section{Proposed Method} \label{Section_Proposed_Method}

In this section, we propose FastTuckerPlus, an SGD-based non-convex optimization algorithm. FastTuckerPlus is a local search algorithm that demonstrates excellent convergence properties. It offers advantages in terms of computational overhead and memory access overhead. Moreover, it exhibits strong adaptability, allowing it to leverage the performance of Tensor Cores effectively. We provide a comprehensive description of the FastTuckerPlus algorithm in Section \ref{Section_A_Non-Convex_SGD-based_Sparse_FastTucker_Decomposition_Algorithm}, highlighting its key details. To optimize the calculation process for Tensor Cores and maximize their performance, we introduce matrixization in Section \ref{Section_Matrixization}. Lastly, we conduct a complexity analysis of FastTuckerPlus in Section \ref{Section_Complexity_Analysis}.

\subsection{A Non-Convex SGD-based Sparse FastTucker Decomposition Algorithm}\label{Section_A_Non-Convex_SGD-based_Sparse_FastTucker_Decomposition_Algorithm}

Our proposed Sparse FastTuckerPlus Decomposition alternately solves non-convex optimization objective
Our proposed FastTuckerPlus Decomposition algorithm alternately solves two non-convex optimization objectives, objective
\begin{equation}\label{low_rank_optimization_mutil}
	\tiny
	\begin{aligned}
		\mathop{\arg\min}_{\textbf{a}^{(n)}_{i_{n},:}, n \in\{N\}}
		f\bigg(\textbf{a}^{(n)}_{i_{n},:}\bigg|x_{i_1,\dots,i_N}, \big\{\textbf{a}^{(n)}_{i_{n},:}\big\}, \big\{\textbf{B}^{(n)}\big\}\bigg)
		=&\bigg\|x_{i_1,\dots,i_N}- \widehat{x}_{i_1,\dots,i_N}\bigg\|_{2}^{2}\\
		&+\lambda_{\textbf{A}}\sum_{n=1}^{N}\|\textbf{a}^{(n)}_{i_{n},:}\|_{2}^{2}
	\end{aligned}
\end{equation}
and objective
\begin{equation}\label{core_tensor_optimization_mutil}
	\tiny
	\begin{aligned}
		\mathop{\arg\min}_{\textbf{B}^{(n)}, n \in\{N\}}
		f\bigg(\textbf{B}^{(n)}\bigg|x_{i_1,\dots,i_N}, \big\{\textbf{a}^{(n)}_{i_{n},:}\big\}, \big\{\textbf{B}^{(n)}\big\}\bigg)
		=&\bigg\|x_{i_1,\dots,i_N}- \widehat{x}_{i_1,\dots,i_N}\bigg\|_{2}^{2}\\
		&+\lambda_{\textbf{B}}\sum_{n=1}^{N}\|\textbf{B}^{(n)}\|_{2}^{2}
	\end{aligned}
\end{equation}
Although the above two optimization objectives are non-convex, they exhibit a relatively simple optimization landscape, allowing them to be effectively solved using local search methods by
\begin{equation}\label{Update_FastTuckerPlus_factor}
	\tiny
	\begin{aligned}
		\left\{
		\begin{aligned}
			\textbf{a}^{(1)}_{i_{1},:}\leftarrow &\textbf{a}^{(1)}_{i_{1},:}+\gamma_{A}\bigg(\big(x_{i_1,\dots,i_N}
			-\widehat{x}_{i_1,\dots,i_N}\big)\textbf{d}^{(1)}_{i_{1},:}\textbf{B}^{(1)^{T}}
			-\lambda_{\textbf{A}}\textbf{a}^{(1)}_{i_{1},:}\bigg)\\
			\cdots\\
			\cdots\\
			\cdots\\
			\textbf{a}^{(N)}_{i_{N},:}\leftarrow &\textbf{a}^{(N)}_{i_{N},:}+\gamma_{A}\bigg(\big(x_{i_1,\dots,i_N}
			-\widehat{x}_{i_1,\dots,i_N}\big)\textbf{d}^{(N)}_{i_{N},:}\textbf{B}^{(N)^{T}}
			-\lambda_{\textbf{A}}\textbf{a}^{(N)}_{i_{N},:}\bigg)\\
		\end{aligned}
		\right.\\
	\end{aligned}
\end{equation}
for optimization objective (\ref{low_rank_optimization_mutil}) and
\begin{equation}\label{Update_FastTuckerPlus_core}
	\tiny
	\begin{aligned}
		\left\{
		\begin{aligned}
			\textbf{B}^{(1)}\leftarrow &\textbf{B}^{(1)}+\gamma_{B}\bigg(\big(x_{i_1,\dots,i_N}
			-\widehat{x}_{i_1,\dots,i_N}\big)\textbf{a}^{(1)^{T}}_{i_{1},:}\textbf{d}^{(1)}_{i_{1},:}
			-\lambda_{\textbf{B}}\textbf{B}^{(1)}\bigg)\\
			\cdots\\
			\cdots\\
			\cdots\\
			\textbf{B}^{(N)}\leftarrow &\textbf{B}^{(N)}+\gamma_{B}\bigg(\big(x_{i_1,\dots,i_N}
			-\widehat{x}_{i_1,\dots,i_N}\big)\textbf{a}^{(N)^{T}}_{i_{N},:}\textbf{d}^{(N)}_{i_{N},:}
			-\lambda_{\textbf{B}}\textbf{B}^{(N)}\bigg)\\
		\end{aligned}
		\right.\\
	\end{aligned}
\end{equation}
for optimization objective (\ref{core_tensor_optimization_mutil}).
For any given element $x_{i_1,\dots,i_N}$ $\in$ $\Omega$, the updates of $\textbf{a}^{(n)}_{i_{n}}$, $n \in \{N\}$ in equation (\ref{Update_FastTuckerPlus_factor}) are independent of each other. Similarly, the updates of $\textbf{B}^{(n)}$, $n \in \{N\}$ in equation (\ref{Update_FastTuckerPlus_core}) are also independent of each other. This implies that only one element needs to be selected for each update. However, the subset $\Psi$ can still be selected from $\Omega$, and the updates can be performed one by one according to the elements in $\Psi$. Furthermore, for a given $x_{i_1,\dots,i_N}$, its corresponding $\textbf{c}^{(n)}_{i_{n}}$, $n \in \{N\}$ are the same. Therefore, the $\textbf{c}^{(n)}_{i_{n}}$, $n \in \{N\}$ can be precalculated to avoid redundant computations when calculating $\textbf{d}^{(n)}_{i_{n},:}$, $n \in \{N\}$. The unified update of $\textbf{a}^{(n)}_{i_{n}}$, $n \in \{N\}$ or $\textbf{B}^{(n)}$, $n \in \{N\}$ ensures that $\textbf{c}^{(n)}_{i_{n}}$, $n \in \{N\}$ follows the updates. 
Therefore, maintaining $\textbf{C}^{(n)}$, $n \in \{N\}$ in memory does not reduce the computational overhead.
The Sparse FastTuckerPlus Decomposition algorithm is described in Algorithm \ref{Algorithm_FastTuckerPlus}.

\begin{algorithm}[hptb]
	\tiny
	\caption{Sparse FastTuckerPlus Decomposition}
	\label{Algorithm_FastTuckerPlus}
	\vspace{.1cm}
	$\textbf{Input}$: Sparse tensor $\bm{\mathcal{X}}$ $\in$ $\mathbb{R}^{I_{1}\times\cdots\times I_{N}}$, 
	ranks $J_{n}$, $n \in \{N\}$ and $R$,
	learning rates $\gamma_{\textbf{A}}$ and $\gamma_{\textbf{B}}$,
	regularization parameters $\lambda_{\textbf{A}}$ and $\lambda_{\textbf{B}}$, and iterations $T$.\\
	$\textbf{Output}$: Factor matrices $\textbf{A}^{(n)}$, $n$ $\in$ $\{N\}$ and core matrices $\textbf{B}^{(n)}$, $n$ $\in$ $\{N\}$.\\
	\begin{algorithmic}[1]
		\STATE Initialize factor matrices $\textbf{A}^{(n)}$ $\in$ $\mathbb{R}^{I_{n}\times J_{n}}$, $n \in \{N\}$ and core
		matrices $\textbf{B}^{(n)}$ $\in$ $\mathbb{R}^{J_{n}\times R}$, $n \in \{N\}$.
		\FOR{$t$ from $1$ to $T$}
		\FOR{randomly takes $\Psi$ from $\Omega$}
		\STATE Calculate $\textbf{c}^{(n)}_{i_{n},:},$ $n \in \{N\}$.
		\FOR{$n$ from $1$ to $N$}
		\STATE Update $\textbf{a}^{(n)}_{i_{n},:}$ by rule (\ref{Update_FastTuckerPlus_factor}).
		\ENDFOR
		\ENDFOR
		\FOR{randomly takes $\Psi$ from $\Omega$}
		\STATE Calculate $\textbf{c}^{(n)}_{i_{n},:},$ $n \in \{N\}$.
		\FOR{$n$ from $1$ to $N$}
		\STATE Update $\textbf{B}^{(n)}$ by rule (\ref{Update_FastTuckerPlus_core}).
		\ENDFOR
		\ENDFOR	
		\ENDFOR		
	\end{algorithmic}
\end{algorithm}

\subsection{Matrixization} \label{Section_Matrixization}

To leverage the computational capabilities of Tensor Cores, we can express the update rules (\ref{Update_FastTuckerPlus_factor}) and (\ref{Update_FastTuckerPlus_core}) in the form of matrix calculations. We randomly select a subset $\Psi$ from $\Omega$ containing $M$ elements and represent it as a matrix $\textbf{X}_{\Psi}$ $\in$ $\mathbb{R}^{M \times 1}$, where $x_{m,1}$ corresponds to the $m$-th element in $\Psi$. We use $\Psi^{(n)}$ to denote the index set of elements in $\Psi$ for the $n$-th order, and $\Psi_{i}^{(n)}$ to denote the $i$-th index in $\Psi^{(n)}$.
To form the factor sub-matrix $\textbf{A}_{\Psi^{(n)},:}^{(n)}$, we extract the factor vector with index set $\Psi^{(n)}$ from the factor matrix $\textbf{A}^{(n)}$. Then, we define $\textbf{C}_{\Psi^{(n)},:}^{(n)}$ $=$ $\textbf{A}_{\Psi^{(n)},:}^{(n)}$ $\textbf{B}^{(n)}$ $\in$ $\mathbb{R}^{M \times R}$, where the $(\Psi_{i}^{(n)}, r)$-th element $c_{\Psi_{i}^{(n)}, r}^{(n)}$ $=$ $\textbf{a}^{(n)}_{\Psi_{i}^{(n)}, r} \cdot \textbf{b}^{(n)}_{:,r}$ of $\textbf{C}_{\Psi^{(n)},:}^{(n)}$.
Similarly to $\textbf{d}^{(n)}_{i_{n},:}$, we use $\textbf{D}_{\Psi^{(n)},:}^{(n)}$ $\in$ $\mathbb{R}^{M \times R}$ to denote $\textbf{C}_{\Psi^{(1)},:}^{(1)}$ $*$ $\cdots$ $*$ $\textbf{C}_{\Psi^{(n-1)},:}^{(n-1)}$ $*$ $\textbf{C}_{\Psi^{(n+1)},:}^{(n+1)}$ $*$ $\cdots$ $*$ $\textbf{C}_{\Psi^{(N)},:}^{(N)}$.

Then update rules (\ref{Update_FastTuckerPlus_factor}) and (\ref{Update_FastTuckerPlus_core}) can be expressed as follows:
\begin{equation}\label{Update_FastTuckerPlus_Matrix_factor}
	\tiny
	\begin{aligned}
		\left\{
		\begin{aligned}
			\textbf{A}_{\Psi^{(1)}}^{(1)}\leftarrow&\textbf{A}_{\Psi^{(1)},:}^{(1)}+\gamma_{A}\bigg((\textbf{X}_{\Psi}
			-\widehat{\textbf{X}}_{\Psi})\circledast(\textbf{D}_{\Psi^{(1)},:}^{(1)}\textbf{B}^{(1)^{T}})
			-\lambda_{\textbf{A}}\textbf{A}_{\Psi^{(1)},:}^{(1)}\bigg)\\
			\cdots\\
			\cdots\\
			\cdots\\
			\textbf{A}_{\Psi^{(N)},:}^{(N)}\leftarrow&\textbf{A}_{\Psi^{(N)},:}^{(N)}+\gamma_{A}\bigg((\textbf{X}_{\Psi}
			-\widehat{\textbf{X}}_{\Psi})\circledast(\textbf{D}_{\Psi^{(N)},:}^{(N)}\textbf{B}^{(N)^{T}})
			-\lambda_{\textbf{A}}\textbf{A}_{\Psi^{(N)},:}^{(N)}\bigg)\\
		\end{aligned}
		\right.\\
	\end{aligned}
\end{equation}
where $\widehat{\textbf{X}}_{\Psi}$ $=$ $\textbf{A}_{\Psi^{(1)},:}^{(1)}\odot(\textbf{B}^{(1)}\textbf{D}_{\Psi^{(1)},:}^{(1)^{T}})$ $\in$ $\mathbb{R}^{M \times 1}$, and
\begin{equation}\label{Update_FastTuckerPlus_Matrix_core}
	\tiny
	\begin{aligned}
		\left\{
		\begin{aligned}
			\textbf{B}^{(1)}\leftarrow &\textbf{B}^{(1)}+\gamma_{B}\bigg(\big((\textbf{X}_{\Psi}
			-\widehat{\textbf{X}}_{\Psi})\circledast\textbf{A}_{\Psi^{(1)},:}^{(1)}\big)^{T}\textbf{D}_{\Psi^{(1)},:}^{(1)}
			-\lambda_{\textbf{B}}\textbf{B}^{(1)}\bigg)\\
			\cdots\\
			\cdots\\
			\cdots\\
			\textbf{B}^{(N)}\leftarrow &\textbf{B}^{(N)}+\gamma_{B}\bigg(\big((\textbf{X}_{\Psi}
			-\widehat{\textbf{X}}_{\Psi})\circledast\textbf{A}_{\Psi^{(N)},:}^{(N)}\big)^{T}\textbf{D}_{\Psi^{(N)},:}^{(N)}
			-\lambda_{\textbf{B}}\textbf{B}^{(N)}\bigg)\\
		\end{aligned}
		\right.\\
	\end{aligned}
\end{equation}
where $\widehat{\textbf{X}}_{\Psi}$ $=$ $\textbf{C}_{\Psi^{(1)},:}^{(1)}\odot\textbf{D}_{\Psi^{(1)},:}^{(1)^{T}}$ $\in$ $\mathbb{R}^{M \times 1}$.
And we use $\textbf{E}_{\Psi^{(n)},:}^{(n)}$ $\in$ $\mathbb{R}^{M \times J_n}$ to denote $(\textbf{X}_{\Psi}
-\widehat{\textbf{X}}_{\Psi})\circledast\textbf{A}_{\Psi^{(N)},:}^{(N)}$.
The matrix calculation process that can be accelerated by Tensor Cores is as follows:
$\textbf{A}_{\Psi^{(n)},:}^{(n)}$ $\cdot$ $\textbf{B}^{(n)}$, $n$ $\in$ $\{N\}$, 
$\textbf{D}_{\Psi^{(n)},:}^{(n)}$ $\cdot$ $\textbf{B}^{(n)^{T}}$, $n$ $\in$ $\{N\}$
and $\textbf{E}_{\Psi^{(n)},:}^{(n)^{T}}$ $\cdot$ $\textbf{D}_{\Psi^{(n)},:}^{(n)}$, $n$ $\in$ $\{N\}$.

Unlike FastTuckerPlus Decomposition, which samples from $\Omega$ completely randomly, FastTucker Decomposition and FasterTucker Decomposition adhere to different constraints. Table \ref{Sample} describes the differences in sampling between the FastTuckerPlus Decomposition algorithm and the SOTA algorithms. The FastTuckerPlus Decomposition algorithm offers greater flexibility in terms of sampling methods.
\begin{table*}[htbp]
	\caption{The sampling of FastTuckerPlus Decomposition and SOTA algorithms.}
	\footnotesize
	\centering
	\label{Sample}
	\begin{tabular}{c|ccc}
		\hline
		\hline
		Algorithms    	& Sample set                       & Sample set source       
		& Description of sample set source   
		\\
		\hline
		FastTucker    	& $\Psi$ for $i_n$ and $n$         & $\Omega_{i_{n}}^{(n)}$  
		& \makecell[c]{The set of non-zero elements\\ whose $n$th order index is $i_{n}$.} 
		\\	
		FasterTucker  	& $\Psi$ for $i_n$ and $n$         & $\Omega_{i_1,\cdots,i_{n-1},i_{n+1},\cdots,i_N}^{(n)}$ 
		& \makecell[c]{The set of non-zero elements whose\\ $(1, \cdots, n-1, n+1, \cdots, N)$th order index\\ are $(i_{1}, \cdots, i_{n-1}, i_{n+1}, \cdots, i_{N})$.}    
		\\
		FastTuckerPlus	& $\Psi$ for all $i_n$ and all $n$ & $\Omega$ 
		& The whole $\Omega$.        	   
		\\
		\hline
		\hline
	\end{tabular}
\end{table*}

It is worth mentioning that the update rules of Algorithm \ref{Algorithm_FastTucker} for Sparse FastTucker Decomposition, as defined by Equations (\ref{SGD}), (\ref{Gradient_low_rank_cp}), and (\ref{Gradient_core_tensor_cp}), can be expressed in matrix form as follows:
\begin{equation}\label{update_fasttucker_factor}
	\tiny
	\begin{aligned}
		\textbf{a}_{i_n,:}^{(n)}\leftarrow\textbf{a}_{i_n,:}^{(n)}+\gamma_{\textbf{A}}\bigg(\big(\textbf{X}_{i_n,:}^{(n)}
		-\textbf{a}^{(n)}_{i_{n},:}\textbf{B}^{(n)}\textbf{D}_{\Psi^{(n)},:}^{(n)^{T}}\big)\textbf{D}_{\Psi^{(n)},:}^{(n)}\textbf{B}^{(n)^{T}}
		-\lambda_{\textbf{A}}\textbf{a}_{i_n,:}^{(n)}\bigg)
	\end{aligned}
\end{equation}
for each $\textbf{a}_{i_n,:}^{(n)}$, $i_n \in \{I_n\}$, $n \in \{N\}$ and
\begin{equation}\label{update_fasttucker_core}
	\tiny
	\begin{aligned}
		\textbf{B}^{(n)}\leftarrow \textbf{B}^{(n)}+\gamma_{\textbf{B}}\bigg(\textbf{a}_{i_n,:}^{(n)^{T}}\big(\textbf{X}_{i_n,:}^{(n)}
		-\textbf{c}^{(n)}_{i_{n},:}\textbf{D}_{\Psi^{(n)},:}^{(n)^{T}}\big)\textbf{D}_{\Psi^{(n)},:}^{(n)}
		-\lambda_{\textbf{B}}\textbf{B}^{(n)}\bigg)
	\end{aligned}
\end{equation}
for each $\textbf{B}^{(n)}$, $n \in \{N\}$.
In the above update process, 
the operations $\textbf{A}_{\Psi^{(n)},:}^{(n)}$ $\cdot$ $\textbf{B}^{(n)}$, $n$ $\in$ $\{N\}$, 
$\textbf{D}_{\Psi^{(n)},:}^{(n)}$ $\cdot$ $\textbf{B}^{(n)^{T}}$, $n$ $\in$ $\{N\}$
and $\bigg(\textbf{a}_{i_n,:}^{(n)^{T}}\big(\textbf{X}_{i_n,:}^{(n)}
-\textbf{c}^{(n)}_{i_{n},:}  \textbf{D}_{\Psi^{(n)},:}^{(n)^{T}}\big)\bigg)\cdot\textbf{D}_{\Psi^{(n)},:}^{(n)}$, $n$ $\in$ $\{N\}$
can be accelerated by Tensor Cores.

Also according to Equations (\ref{SGD}), (\ref{Gradient_low_rank_cp}) and (\ref{Gradient_core_tensor_cp}), the update rules of Algorithm \ref{Algorithm_FasterTucker} for Sparse FasterTucker Decomposition can be expressed as matrix operations:
\begin{equation}\label{update_fastertucker_factor}
	\tiny
	\begin{aligned}
		\textbf{A}_{\Psi^{(n)},:}^{(n)}\leftarrow\textbf{A}_{\Psi^{(n)},:}^{(n)}+\gamma_{\textbf{A}}\bigg(\big(\textbf{X}_{:,\Psi^{(n)}}^{(n)^{T}}
		-\textbf{A}^{(n)}_{\Psi^{(n)},:}\textbf{B}^{(n)}\textbf{d}_{i_n,:}^{(n)^{T}}\big)\textbf{d}_{i_n,:}^{(n)}\textbf{B}^{(n)^{T}}
		-\lambda_{\textbf{A}}\textbf{A}_{\Psi^{(n)},:}^{(n)}\bigg)
	\end{aligned}
\end{equation}
for each $\textbf{a}_{i_n,:}^{(n)}$, $i_n \in \{I_n\}$, $n \in \{N\}$ and
\begin{equation}\label{update_fastertucker_core}
	\tiny
	\begin{aligned}
		\textbf{B}^{(n)}\leftarrow \textbf{B}^{(n)}+\gamma_{\textbf{B}}\bigg(\textbf{A}^{(n)^{T}}_{\Psi^{(n)},:}\big(\textbf{X}_{:,\Psi^{(n)}}^{(n)^{T}}
		-\textbf{C}^{(n)}_{\Psi^{(n)},:}\textbf{d}_{i_n,:}^{(n)^{T}}\big)\textbf{d}_{i_n,:}^{(n)}
		-\lambda_{\textbf{B}}\textbf{B}^{(n)}\bigg)
	\end{aligned}
\end{equation}
for each $\textbf{B}^{(n)}$, $n \in \{N\}$.

Prior to the aforementioned process, $\textbf{C}^{(n)}$, $n \in \{N\}$ has been pre-calculated and stored in memory. Therefore, only the operation $\textbf{A}^{(n)^{T}}_{\Psi^{(n)},:}\cdot\bigg(\big(\textbf{X}_{:,\Psi^{(n)}}^{(n)^{T}}
-\textbf{C}^{(n)}_{\Psi^{(n)},:}\textbf{d}_{i_n,:}^{(n)^{T}}\big)\textbf{d}_{i_n,:}^{(n)}\bigg)$, $n \in \{N\}$ can be accelerated using Tensor Cores. However, multiple different $\textbf{d}_{i_n,:}^{(n)}$ from various sample sets $\Psi$ can be combined to form a larger matrix $\textbf{D}_{i_n,:}^{(n)}$, which allows the operation $\textbf{B}^{(n)}\textbf{d}_{i_n,:}^{(n)^{T}}$ in $\textbf{B}^{(n)}\textbf{D}_{i_n,:}^{(n)^{T}}$ to be accelerated using Tensor Cores. In our experiments, we have implemented cuFastTuckerTC, cuFasterTucker\_TC, and cuFasterTuckerCOO\_TC, which utilize Tensor Cores to accelerate cuFastTucker, cuFasterTucker, and cuFasterTuckerCOO, respectively.

\subsection{Complexity Analysis} \label{Section_Complexity_Analysis}

\begin{table*}[htbp]
	\caption{The complexity of FastTuckerPlus Decomposition algorithm and SOTA algorithms.}
	\centering
	\footnotesize
	\label{Complexity}
	\begin{tabular}{c|ccc}
		\hline
		\hline
		~&\makecell[c]{FastTucker\\Decomposition}    
		& \makecell[c]{FasterTucker\\Decomposition}
		& \makecell[c]{FastTuckerPlus\\Decomposition}
		\\
		\hline
		\makecell[c]{Read from memory}
		& \makecell[c]{$\textbf{a}_{i_n,:}^{(n)}$\\
			$\textbf{A}_{\Psi^{(k)},:}^{(k)}$, $k \neq n$, $k \in \{N\}$\\
			$\textbf{B}^{(k)}$, $k \in \{N\}$}    
		& \makecell[c]{$\textbf{A}_{\Psi^{(n)},:}^{(n)}$\\
			$\textbf{c}_{i_k,:}^{(k)}$, $k \neq n$, $k \in \{N\}$\\
			$\textbf{B}^{(n)}$} 
		& \makecell[c]{$\textbf{A}_{\Psi^{(n)},:}^{(n)}$, $n \in \{N\}$\\
			$\textbf{B}^{(n)}$, $n \in \{N\}$} 
		\\
		\cline{2-4}
		\makecell[c]{Complexity} 
		& \makecell[c]{$J_n$\\
			$M\sum_{k=1, k\neq n}^{N}J_k$\\
			$R\sum_{n=1}^{N}J_k$}    
		& \makecell[c]{$MJ_n$\\
			$(N-1)R$\\
			$RJ_n$} 
		& \makecell[c]{$M\sum_{n=1}^{N}J_n$\\
			$R\sum_{n=1}^{N}J_n$} \\
		\cline{2-4}
		\makecell[c]{Total for all $n$} 
		& \makecell[c]{$(MN-M+R+1)\sum_{n=1}^{N}J_n$ }    
		& \makecell[c]{$(M+R)\sum_{n=1}^{N}J_n$$+N(N-1)R$}
		& \makecell[c]{$(M+R)\sum_{n=1}^{N}J_n$}
		\\
		\hline
		\makecell[c]{Calculation} 
		& \makecell[c]{$\textbf{D}_{\Psi^{(n)},:}^{(n)}$}
		& \makecell[c]{$\textbf{d}_{i_n,:}^{(n)}$}
		& \makecell[c]{$\textbf{D}_{\Psi^{(n)},:}^{(n)}, n \in \{N\}$}
		\\	
		\makecell[c]{Complexity} 
		& \makecell[c]{$MR(\sum_{k=1, k \neq n}^{N}J_k+N-2)$}
		& \makecell[c]{$(N-2)R$}
		& \makecell[c]{$MR\big(\sum_{n=1}^{N}J_n+N(N-2)\big)$}
		\\	
		\cline{2-4}
		\makecell[c]{Total for all $n$}
		& \makecell[c]{$MR\big((N-1)\sum_{n=1}^{N}J_n+N(N-2)\big)$}
		& \makecell[c]{$N(N-2)R$}
		& \makecell[c]{$MR\big(\sum_{n=1}^{N}J_n+N(N-2)\big)$}
		\\
		\hline
		\makecell[c]{Calculation} 
		& \makecell[c]{$\textbf{B}^{(n)}\textbf{D}_{\Psi^{(n)},:}^{(n)^{T}}$}
		& \makecell[c]{$\textbf{B}^{(n)}\textbf{d}_{i_n,:}^{(n)^{T}}$}
		& \makecell[c]{$\textbf{B}^{(n)}\textbf{D}_{\Psi^{(n)},:}^{(n)^{T}}, n \in \{N\}$}
		\\
		\makecell[c]{Complexity} 
		& \makecell[c]{$MRJ_n$}
		& \makecell[c]{$RJ_n$}
		& \makecell[c]{$MR\sum_{n=1}^{N}J_n$}
		\\	
		\cline{2-4}
		\makecell[c]{Total for all $n$}
		& \makecell[c]{$MR\sum_{n=1}^{N}J_n$}
		& \makecell[c]{$R\sum_{n=1}^{N}J_n$}
		& \makecell[c]{$MR\sum_{n=1}^{N}J_n$}
		\\
		\hline
		\makecell[c]{Update} 
		& \makecell[c]{$\textbf{a}_{i_n,:}^{(n)}$}
		& \makecell[c]{$\textbf{A}_{\Psi^{(n)},:}^{(n)}$}
		& \makecell[c]{$\textbf{A}_{\Psi^{(n)},:}^{(n)}$}
		\\
		\makecell[c]{Complexity} 
		& \makecell[c]{$J_n$}
		& \makecell[c]{$MJ_n$}
		& \makecell[c]{$M\sum_{n=1}^{N}J_n$}
		\\	
		\cline{2-4}
		\makecell[c]{Total for all $n$} &\makecell[c]{$\sum_{n=1}^{N}J_n$}
		&
		\makecell[c]{$M\sum_{n=1}^{N}J_n$}
		&
		\makecell[c]{$M\sum_{n=1}^{N}J_n$}
		\\
		\hline
		\hline
	\end{tabular}
\end{table*}

In Algorithm \ref{Algorithm_FastTucker}, when updating $\textbf{a}^{(n)}_{i_{n}}$ or $\textbf{B}^{(n)}$ using a single $\Psi$ with $M$ selected samples, the following data need to be read from memory: a matrix set consisting of $N$ matrices of size $J_{k} \times R$, $k$ $\in$ $\{N\}$ $\big($$\textbf{B}^{(k)}$, $k$ $\in$ $\{N\}$$\big)$, a matrix set consisting of $N-1$ matrices of size $M \times J_{k}$ $\big($$\textbf{A}_{\Psi^{(k)},:}^{(k)}$, $k \neq n$, $k \in \{N\}\big)$, and a vector of length $J_n$ $\big($ $\textbf{a}^{(n)}_{i_{n},:}$$\big)$. For all $n$ $\in$ $\{N\}$, a total of $(MN-M+R+1)\sum_{n=1}^{N}J_n$ parameters are read from memory.
In Algorithm \ref{Algorithm_FasterTucker}, when updating $\textbf{a}^{(n)}_{i_{n}}$ or $\textbf{B}^{(n)}$ using a single $\Psi$ with $M$ selected samples, the following data need to be read from memory: a matrix of size $J_{n} \times R$ $\big($$\textbf{B}^{(n)}$$\big)$, a vector set consisting of $N-1$ vectors of length $R$ $\big($$\textbf{c}_{i_k,:}^{(k)}$, $k \neq n$, $k \in \{N\}\big)$, and a matrix of size $M \times J_{n}$ $\big($ $\textbf{A}_{\Psi^{(n)},:}^{(n)}$$\big)$. For all $n$ $\in$ $\{N\}$, a total of $(M+R)\sum_{n=1}^{N}J_n+N(N-1)R$ parameters are read from memory.
However, in practice, most $\Omega_{i_1,\cdots,i_{n-1},i_{n+1},\cdots,i_N}^{(n)}$ contain fewer than $M$ elements, resulting in $\Psi$ containing fewer than $M$ elements.
In Algorithm \ref{Algorithm_FastTuckerPlus}, when updating $\textbf{a}^{(n)}_{i_{n}}$, $n$ $\in$ $\{N\}$ or $\textbf{B}^{(n)}$, $n$ $\in$ $\{N\}$ using a single $\Psi$ with $M$ selected samples, the following data need to be read from memory: a matrix set consisting of $N$ matrices of size $J_{n} \times R$ $\big($$\textbf{B}^{(n)}$, $n$ $\in$ $\{N\}$$\big)$ and a matrix set consisting of $N$ matrices of size $M \times J_{n}$ $\big($$\textbf{A}_{\Psi^{(n)},:}^{(n)}$, $n \in \{N\}$$\big)$. A total of  $(M+R)\sum_{n=1}^{N}J_n$ parameters are read from memory.
Throughout the entire process, FastTuckerPlus reads fewer parameters from memory compared to FastTucker and FasterTucker.

The main difference in terms of computational complexity among algorithms \ref{Algorithm_FastTucker}, \ref{Algorithm_FasterTucker}, and \ref{Algorithm_FastTuckerPlus} lies in the calculation method of $\textbf{c}^{(n)}_{i_{n}}$, $i_n \in \{I_n\}$, $n \in \{N\}$.
In algorithm \ref{Algorithm_FastTucker}, when updating $\textbf{a}^{(n)}_{i_{n}}$ or $\textbf{B}^{(n)}$ for the entire $\Omega$, it requires $|\Omega|\sum_{k=1, k\neq n}^{N}J_{k}R$ multiplications. For all $n$ $\in$ $\{N\}$, the total number of multiplications is $|\Omega|(N-1)\sum_{n=1}^{N}J_{n}R$.
In algorithm \ref{Algorithm_FasterTucker}, when updating $\textbf{a}^{(n)}_{i_{n}}$ or $\textbf{B}^{(n)}$ for the entire $\Omega$, it only requires $I_{n}J_{n}R$ multiplications. For all $n$ $\in$ $\{N\}$, the total number of multiplications is $\sum_{n=1}^{N}I_nJ_{n}R$.
In algorithm \ref{Algorithm_FastTuckerPlus}, when updating $\textbf{a}^{(n)}_{i_{n}}$, $n$ $\in$ $\{N\}$ or $\textbf{B}^{(n)}$, $n$ $\in$ $\{N\}$ for the entire $\Omega$, it requires $|\Omega|\sum_{n=1}^{N}J_{n}R$ multiplications.
Since in sparse tensors, $I_n, n \in \{N\}$ is much smaller than $|\Omega|$, we have $\sum_{n=1}^{N}I_nJ_{n}R$ $\ll$ $|\Omega|\sum_{n=1}^{N}J_{n}R$ $<$ $|\Omega|(N-1)\sum_{n=1}^{N}J_{n}R$.
Based on this observation, in the subsequent analysis, the computational complexity of calculating $\textbf{c}^{(n)}_{i_{n}}$, $i_n \in \{I_n\}$, $n \in \{N\}$ in Algorithm \ref{Algorithm_FasterTucker} is ignored.

The computation of $\textbf{D}_{\Psi^{(n)},:}^{(n)}, n \in \{N\}$ or $\textbf{d}_{i_n,:}^{(n)}, n \in \{N\}$ is a crucial step in these algorithms. In Algorithm \ref{Algorithm_FastTucker}, when updating $\textbf{a}^{(n)}_{i_{n}}$ or $\textbf{B}^{(n)}$ with a single $\Psi$ containing $M$ selected samples, the number of required multiplications to compute $\textbf{D}_{\Psi^{(n)},:}^{(n)}$ is $MR(\sum_{k=1, k \neq n}^{N}J_k+N-2)$. For all $n$ $\in$ $\{N\}$, the total number of multiplications is $MR\big((N-1)\sum_{n=1}^{N}J_n+N(N-2)\big)$.
In Algorithm \ref{Algorithm_FasterTucker}, the $\textbf{C}^{(k)}$, $k \neq n$, $k \in \{N\}$, have already been calculated and stored in memory for updating $\textbf{a}^{(n)}_{i_{n}}$ or $\textbf{B}^{(n)}$ using a single $\Psi$ containing $M$ samples. The number of required multiplications to compute $\textbf{d}_{i_n,:}^{(n)}$ is $(N-2)R$. For all $n$ $\in$ $\{N\}$, the total number of multiplications is $N(N-2)R$.
In Algorithm \ref{Algorithm_FastTuckerPlus}, when updating $\textbf{a}^{(n)}_{i_{n}}$, $n$ $\in$ $\{N\}$ or $\textbf{B}^{(n)}$, $n$ $\in$ $\{N\}$ using a single $\Psi$ containing $M$ samples, after calculating all $\textbf{C}_{\Psi^{(n)},:}^{(n)}, n \in \{N\}$, they are shared by all $\textbf{D}_{\Psi^{(n)},:}^{(n)}, n \in \{N\}$ during the computation of $\textbf{D}_{\Psi^{(n)},:}^{(n)}, n \in \{N\}$. Hence, the total number of required multiplications in the entire process is $MR\big(\sum_{n=1}^{N}J_n+N(N-2)\big)$.

Algorithm \ref{Algorithm_FasterTucker} offers the advantage of only requiring the computation of $\textbf{B}^{(n)}\textbf{d}_{i_n,:}^{(n)^{T}}, n \in \{N\}$, instead of $\textbf{B}^{(n)}\textbf{D}_{\Psi^{(n)},:}^{(n)^{T}}, n \in \{N\}$. This leads to a reduced computational workload. On the other hand, Algorithm \ref{Algorithm_FastTucker} has the advantage of updating fewer parameters, resulting in lower complexity when writing back to memory.
However, both algorithms encounter the challenge of load imbalance in the sampling process when applied to sparse tensors.

Table \ref{Complexity} provides a summary of the analysis discussed above. FastTuckerPlus Decomposition, compared to FastTucker Decomposition, combines $2N$ update processes into $2$ update processes, resulting in a reduction in the reading process of $\textbf{A}_{\Psi^{(n)},:}^{(n)}$, $n$ $\in$ $\{N\}$ and $\textbf{B}^{(n)}$, $n$ $\in$ $\{N\}$ from memory. Faster Decomposition converts the memory reading process of $\textbf{a}{i_k,:}^{(k)}$ for $k \neq n$, $k \in {N}$ and $\textbf{B}^{(k)}$ for $k \neq n$, $k \in {N}$ into the reading process of $\textbf{c}_{i_k,:}^{(k)}$, $k \neq n$, $k \in \{N\}$, thereby reducing the number of parameters to be read and the computational burden of $\textbf{a}^{(k)}_{i_{k}}$$\cdot$$\textbf{B}^{(k)}$, $k \neq n$, $k$ $\in$ $\{N\}$. This optimization is possible because when updating $\textbf{A}_{\Psi^{(n)},:}^{(n)}$ or $\textbf{B}^{(n)}$, the values of $\textbf{A}_{\Psi^{(k)},:}^{(k)}$, $k \neq n$, $k$ $\in$ $\{N\}$ and $\textbf{B}^{(k)}$, $k \neq n$, $k$ $\in$ $\{N\}$ remain unchanged. However, in FastTuckerPlus, all $\textbf{A}_{\Psi^{(n)},:}^{(n)}$, $n$ $\in$ $\{N\}$ or $\textbf{B}^{(n)}$, $n$ $\in$ $\{N\}$ are updated, which requires the real-time computation of $\textbf{D}_{\Psi^{(n)},:}^{(n)}, n \in \{N\}$ during the update process. FasterTucker reduces the computational burden in this aspect by reading $N(N-2)R$ additional parameters from memory compared to FastTuckerPlus. In the experimental section, we will demonstrate that when utilizing Tensor Cores for this computation, it takes less time than reading from memory.

\section{cuFastTuckerPlus On GPU} \label{Section_cuFastTuckerPlus_On_GPU}
We present the implementation of our proposed cuFastTuckerPlus algorithm on GPUs with Tensor Cores. In Section \ref{Section_Tensor_Core}, we provide a brief introduction to Tensor Cores. We also explain how to optimize data partitioning to maximize the utilization of Tensor Cores in Section \ref{Section_Matrix_partition}. Our algorithm is designed to leverage two levels of parallelism: warp parallelism, discussed in Section \ref{Section_Warp_Parallelization}, and block parallelism, explained in Section \ref{Section_Block_Parallelization}. Finally, in Section \ref{Section_Overview}, we summarize our algorithm and highlight the key techniques used in its implementation.

\subsection{Tensor Core} \label{Section_Tensor_Core}

Tensor Cores are specialized processing cores in NVIDIA GPUs that excel in performing matrix operations. Unlike CUDA Cores, which are more general-purpose, Tensor Cores are specifically designed for high-performance matrix computations. Utilizing Tensor Cores can significantly enhance the peak throughput compared to using CUDA Cores for matrix operations.
Tensor Cores perform a fused multiply-add operation. They multiply two $4\times4$ half-precision float matrices, add the result to a $4\times4$ half-precision or single-precision matrix, and produce a new $4\times4$ half-precision or single-precision matrix. NVIDIA refers to these operations performed by Tensor Cores as mixed-precision math because the input matrices are in half-precision, but the product can be in full-precision.
CUDA provides the Warp-level Matrix Multiply and Accumulate (WMMA) API, which allows developers to leverage Tensor Cores on the GPU. Through the WMMA API, developers can calculate $\textbf{D} = \textbf{A} \cdot \textbf{B} + \textbf{C}$ within a warp, where $\textbf{A}$, $\textbf{B}$, $\textbf{C}$, and $\textbf{D}$ can also be tiles of larger matrices. All threads within the warp cooperate to perform their respective matrix multiply-add operations.
The size limit of matrices in WMMA is $M \times N \times K$, where $\textbf{A} \in \mathbb{R}^{M\times K}$, $\textbf{B} \in \mathbb{R}^{N\times K}$, $\textbf{C} \in \mathbb{R}^{M\times N}$, and $\textbf{D} \in \mathbb{R}^{M\times N}$. Currently, CUDA supports $16\times16\times16$, $32\times8\times16$, and $8\times32\times16$ half-precision, as well as $16\times16\times8$ single-precision, and $8\times8\times4$ double-precision matrix multiply-add operations.
For larger matrices, the matrix multiply-add operations can be divided into multiple tile-wise operations of suitable sizes. In our paper, without loss of generality, we utilize $16\times16\times16$ half-precision matrix multiply-add operations.

\subsection{Matrix partition} \label{Section_Matrix_partition}

To maximize performance and efficiency, we divide $\textbf{A}_{\Psi^{(n)},:}^{(n)}$, $n$ $\in$ $\{N\}$ and $\textbf{B}^{(n)}$, $n$ $\in$ $\{N\}$ into multiple tiles of submatrices with a size of $16\times16$. Any remaining entries are padded with zeros. It is worth noting that the best performance can be achieved when $M$, $R$ and $J_n$, $n$ $\in$ $\{N\}$ are all multiples of $16$.
For the sake of convenience and to save memory costs, we set $M$ to be $16$. Let $J_n = 16P_n$, $n$ $\in$ $\{N\}$, and $R = 16Q$. As a result, $\textbf{A}_{\Psi^{(n)},:}^{(n)}$, $n$ $\in$ $\{N\}$ can be divided into $1\times P_n$ tiles, and $\textbf{B}^{(n)}$, $n$ $\in$ $\{N\}$ can be divided into $P_n\times Q$, $n$ $\in$ $\{N\}$ tiles. Correspondingly, $\textbf{D}_{\Psi^{(n)}}^{(n)}$, $n$ $\in$ $\{N\}$, are divided into $1\times Q$ tiles, and $\textbf{E}_{\Psi^{(n)}}^{(n)^{T}}$, $n$ $\in$ $\{N\}$ are divided into $1\times P_n$, $n$ $\in$ $\{N\}$ tiles, respectively.

In our notation, $(\textbf{A}_{\Psi^{(n)},:}^{(n)})_{1,p_n}$, represents the $(1,p_n)$-th tile of $\textbf{A}_{\Psi^{(n)},:}^{(n)}$ for $n$ $\in$ $\{N\}$. Similarly, $(\textbf{B}^{(n)})_{p_n,q}$ represents the $(p_n,q)$-th tile of $\textbf{B}^{(n)}$ for $n$ $\in$ $\{N\}$. The computation of $(\textbf{A}_{\Psi^{(n)},:}^{(n)}\textbf{B}^{(n)})_{1,q}$ is given by $\sum_{p_n=1}^{P_n}(\textbf{A}_{\Psi^{(n)},:}^{(n)})_{1,p_n}(\textbf{B}^{(n)})_{p_n,q}$, where $(\textbf{A}_{\Psi^{(n)},:}^{(n)}\textbf{B}^{(n)})_{1,q}$ represents the $(1,q)$-th tile of the matrix product $\textbf{A}_{\Psi^{(n)},:}^{(n)}\textbf{B}^{(n)}$ for $n$ $\in$ $\{N\}$.
Similarly, the computations $\textbf{D}_{\Psi^{(n)},:}^{(n)}$ $\cdot$ $\textbf{B}^{(n)^{T}}$, $n$ $\in$ $\{N\}$ and $\textbf{E}_{\Psi^{(n)},:}^{(n)^{T}}$ $\cdot$ $\textbf{B}^{(n)}$, $n$ $\in$ $\{N\}$ follow the same pattern. Please note that in the paper, we have omitted explicit mention of matrix partitioning for brevity and clarity of the expressions.

\subsection{Warp Parallelization} \label{Section_Warp_Parallelization}

The warp serves as the scheduling unit in NVIDIA GPUs, with all threads in a warp executing the same instructions. In the current mainstream NVIDIA GPU architecture, a warp consists of $32$ threads. The GPU assigns $32$ consecutive threads from a block to form a warp. Even if the number of remaining threads is less than $32$, they are still organized into a warp, leaving the excess threads idle. Therefore, proper task distribution is crucial to avoid idle threads within a warp.
In the cuFastTuckerPlus framework, we utilize a warp to process a specific $\Psi$. When updating the factor matrices, the warp updates the $\textbf{A}_{\Psi^{(n)},:}^{(n)}$, $n$ $\in$ $\{N\}$. This warp handles a tile of size $16\times16$ in each iteration. The calculations involving $\textbf{A}_{\Psi^{(n)},:}^{(n)}\textbf{B}^{(n)}$, $n$ $\in$ $\{N\}$ and $\textbf{D}_{\Psi^{(n)},:}^{(n)}$$\textbf{B}^{(n)^{T}}$, $n$ $\in$ $\{N\}$ can be performed using either CUDA Cores or Tensor Cores.
In the case of Tensor Cores, a warp is responsible for multiplying two tiles of size $16\times16$. The warp computes the elements $(\textbf{A}_{\Psi^{(n)},:}^{(n)})_{1,p_n}(\textbf{B}^{(n)})_{p_n,q}$, $p_n$ $\in$ $\{P_n\}$, $q$ $\in$ $\{Q\}$, $n$ $\in$ $\{N\}$, and then computes $(\textbf{A}_{\Psi^{(n)},:}^{(n)}\textbf{B}^{(n)})_{1,q}$ $=$  $\sum_{p_n=1}^{P_n}(\textbf{A}_{\Psi^{(n)},:}^{(n)})_{1,p_n}(\textbf{B}^{(n)})_{p_n,q}$, $q$ $\in$ $\{Q\}$, $n$ $\in$ $\{N\}$. This process completes the calculation of $\textbf{A}_{\Psi^{(n)},:}^{(n)}\textbf{B}^{(n)}$, $n$ $\in$ $\{N\}$. The calculations involving $\textbf{D}_{\Psi^{(n)},:}^{(n)}$$\textbf{B}^{(n)^{T}}$, $n$ $\in$ $\{N\}$ follow a similar pattern.
For CUDA Cores, we divide a warp into two workers, with each worker consisting of 16 threads. When multiplying two tiles of size $16\times16$, one worker calculates the upper half ($8\times16$) of the resulting tile, while the other worker computes the lower half ($8\times16$) of the tile. Other tile operations with a size of $16\times16$, which do not involve matrix multiplication, are also performed using CUDA Cores. Similar to matrix multiplication, each worker is responsible for half of the tile.
When updating the core matrices, the usage of CUDA Cores and Tensor Cores aligns with the update process for the factor matrices. The distinction is that we do not update $\textbf{B}^{(n)}$, $n$ $\in$ $\{N\}$, within the warp. Instead, we accumulate the gradients of $\textbf{B}^{(n)}$, $n$ $\in$ $\{N\}$ for all $\Psi$ in $\Omega$ and subsequently update $\textbf{B}^{(n)}$, $n$ $\in$ $\{N\}$.

\subsection{Block Parallelization} \label{Section_Block_Parallelization}

A block consists of multiple threads that can synchronize and share shared memory, enabling communication between threads. When the number of threads in a block is a multiple of 32, meaning the block consists of multiple warps, all threads can be fully utilized without any idle threads. In cuFastTuckerPlus, since each $\Psi$ can be processed independently and has the same number of elements, we evenly distribute all sampled $\Psi$ to different warps to achieve load balancing.
In hardware, all threads within a block do not execute in parallel simultaneously. Instead, a block is divided into multiple warps allocated to the same Streaming Multiprocessor and queued for execution on the Streaming Processors. The GPU's hardware architecture allows a large number of warps to reside in the SM concurrently. When a warp encounters a wait condition during execution, such as memory read and write delays, the Warp Scheduler swiftly switches to the next eligible warp for execution to maintain high instruction throughput. This architectural design is a fundamental characteristic of GPUs, where numerous warps are employed to hide latency. The abundance of warps enables the GPU to achieve exceptional data throughput.

\subsection{Overview} \label{Section_Overview}

cuFastTuckerPlus is divided into two main parts: updating the factor matrices and updating the core matrices, as described in Algorithm \ref{cuFastTuckerPlus_Factor} and Algorithm \ref{cuFastTuckerPlus_Core}, respectively. In these algorithms, we prioritize storing the parameters in the fastest memory available and maximizing memory reuse. We ensure that shared memory and registers are not excessively occupied, allowing for a sufficient number of active thread blocks during runtime to enhance parallel efficiency. Furthermore, we strive to minimize unnecessary calculations throughout the computation process. Additionally, cuFastTuckerPlus utilizes a load-balanced sampling method. In summary, the main optimization techniques employed by cuFastTuckerPlus are as follows:

\begin{algorithm}[]
	\caption{Update factor matrices of cuFastTuckerPlus}
	\label{cuFastTuckerPlus_Factor}
	\vspace{.1cm}
	\tiny
	$\mathcal{G}\{parameter\}$: parameters in global memory.\\
	$\mathcal{R}\{parameter\}$: parameters in register memory.\\
	$\textbf{Input}$: Sparse tensor $\bm{\mathcal{X}}$ $\in$ $\mathbb{R}^{I_{1}\times\cdots\times I_{N}}$, 
	initialized factor matrices $\textbf{A}^{(n)}$ $\in$ $\mathbb{R}^{I_{n}\times J_{n}}$, $n \in \{N\}$ and core matrices $\textbf{B}^{(n)}$ $\in$ $\mathbb{R}^{J_{n}\times R}$, $n \in \{N\}$,
	learning rate $\gamma_{\textbf{A}}$, regularization parameter $\lambda_{\textbf{A}}$.\\
	$\textbf{Output}$: Factor matrices $\textbf{A}^{(n)}$, $n$ $\in$ $\{N\}$.\\	
	\begin{algorithmic}[1]
		\FOR{$n$ from $1$ to $N$}
		\STATE $\mathcal{R}\{\textbf{B}^{(n)}\}$ $\leftarrow$ $\mathcal{G}\{\textbf{B}^{(n)}\}$
		\ENDFOR
		\FOR{\emph{Warp Parallelization}}
		\FOR{randomly take $\Psi$ from $\Omega$}
		\FOR{$n$ from $1$ to $N$}
		\STATE $\mathcal{R}\{\textbf{D}^{(n)}_{\Psi^{(n)},:}\}$ $\leftarrow$ $1$
		\ENDFOR
		\FOR{$n$ from $1$ to $N$}
		\STATE $\mathcal{R}\{\textbf{A}^{(n)}_{\Psi^{(n)},:}\}$ $\leftarrow$ $\mathcal{S}\{\textbf{A}^{(n)}_{\Psi^{(n)},:}\}$ $\leftarrow$ $\mathcal{G}\{\textbf{A}^{(n)}_{\Psi^{(n)},:}\}$
		\STATE $\mathcal{S}\{\textbf{C}^{(n)}_{\Psi^{(n)},:}\}$ $\leftarrow$ $\mathcal{R}\{\textbf{C}^{(n)}_{\Psi^{(n)},:}\}$ $\leftarrow$ $\mathcal{R}\{\textbf{A}^{(n)}_{\Psi^{(n)},:}\}$ $\cdot$ $\mathcal{R}\{\textbf{B}^{(n)}\}$ by Tensor Cores
		\FOR{$k$ from $1$ to $N$ and $k$ $\neq$ $n$}
		\STATE $\mathcal{R}\{\textbf{D}^{(k)}_{\Psi^{(k)},:}\}$ $\leftarrow$ $\mathcal{R}\{\textbf{D}^{(k)}_{\Psi^{(k)},:}\}$ $*$ $\mathcal{S}\{\textbf{C}^{(n)}_{\Psi^{(n)},:}\}$
		\ENDFOR
		\ENDFOR
		\FOR{$n$ from $1$ to $N$}
		\STATE $\mathcal{R}\{\textbf{D}^{(n)^{T}}_{\Psi^{(n)},:}\}$ $\leftarrow$ $\mathcal{S}\{\textbf{D}^{(n)}_{\Psi^{(n)},:}\}$ $\leftarrow$ $\mathcal{R}\{\textbf{D}^{(n)}_{\Psi^{(n)},:}\}$
		\STATE $\mathcal{S}\{\textbf{B}^{(n)}\textbf{D}^{(n)^{T}}_{\Psi^{(n)},:}\}$ $\leftarrow$ $\mathcal{R}\{\textbf{B}^{(n)}\textbf{D}^{(n)^{T}}_{\Psi^{(n)},:}\}$ $\leftarrow$ $\mathcal{R}\{\textbf{B}^{(n)}\}$ $\cdot$ $\mathcal{R}\{\textbf{D}^{(n)^{T}}_{\Psi^{(n)},:}\}$ by Tensor Cores
		\ENDFOR
		\STATE $\mathcal{R}\{\widehat{\textbf{X}}_{\Psi}\}$ $\leftarrow$ $\mathcal{S}\{\textbf{A}^{(1)}_{\Psi^{(1)},:}\}$ $\odot$ $\mathcal{S}\{\textbf{B}^{(1)}\textbf{D}^{(1)^{T}}_{\Psi^{(1)},:}\}$ - $\mathcal{G}\{\textbf{X}_{\Psi}\}$
		\FOR{$n$ from $1$ to $N$}	
		\STATE $\mathcal{G}\{\textbf{A}^{(n)}_{\Psi^{(n)},:}\}$ $\leftarrow$ $\mathcal{S}\{\textbf{A}^{(n)}_{\Psi^{(n)},:}\}$ $-$ $\gamma_{\textbf{A}}$ $\cdot$ $\big($ $\mathcal{R}\{\widehat{\textbf{X}}_{\Psi}\}$ $\circledast$ 
		$\mathcal{S}\{\textbf{D}^{(n)}_{\Psi^{(n)},:}\textbf{B}^{(n)^{T}}\}$
		$+$	$\lambda_{\textbf{A}}$ $\cdot$ $\mathcal{S}\{\textbf{A}^{(n)}_{\Psi^{(n)},:}\}$ $\big)$	
		\ENDFOR
		\ENDFOR
		\ENDFOR
	\end{algorithmic}
\end{algorithm}

\begin{algorithm}[]
	\caption{Update core matrices of cuFastTuckerPlus}
	\label{cuFastTuckerPlus_Core}
	\tiny
	\vspace{.1cm}
	$\mathcal{G}\{parameter\}$: parameters in global memory.\\
	$\mathcal{R}\{parameter\}$: parameters in register memory.\\
	$Grad(\textbf{B}^{(n)})$: the gradient of $\textbf{B}^{(n)}$.\\
	$\textbf{Input}$: Sparse tensor $\bm{\mathcal{X}}$ $\in$ $\mathbb{R}^{I_{1}\times\cdots\times I_{N}}$, 
	initialized factor matrices $\textbf{A}^{(n)}$ $\in$ $\mathbb{R}^{I_{n}\times J_{n}}$, $n \in \{N\}$ and core matrices $\textbf{B}^{(n)}$ $\in$ $\mathbb{R}^{J_{n}\times R}$, $n \in \{N\}$,
	learning rate $\gamma_{\textbf{B}}$,
	regularization parameter $\lambda_{\textbf{B}}$.\\
	$\textbf{Output}$: Core matrices $\textbf{B}^{(n)}$, $n$ $\in$ $\{N\}$.\\	
	\begin{algorithmic}[1]
		\FOR{$n$ from $1$ to $N$}
		\STATE $\mathcal{R}\{\textbf{B}^{(n)}\}$ $\leftarrow$ $\mathcal{G}\{\textbf{B}^{(n)}\}$
		\STATE $\mathcal{G}\{Grad(\textbf{B}^{(n)})\}$ $\leftarrow$ $0$
		\ENDFOR	
		\FOR{\emph{Warp Parallelization}}
		\FOR{randomly take $\Psi$ from $\Omega$}		
		\FOR{$n$ from $1$ to $N$}
		\STATE $\mathcal{R}\{Grad(\textbf{B}^{(n)})\}$ $\leftarrow$ $0$
		\STATE $\mathcal{R}\{\textbf{D}^{(n)}_{\Psi^{(n)},:}\}$ $\leftarrow$ $1$
		\ENDFOR	
		\FOR{$n$ from $1$ to $N$}
		\STATE $\mathcal{R}\{\textbf{A}^{(n)^{T}}_{\Psi^{(n)},:}\}$ $\leftarrow$ $\mathcal{S}\{\textbf{A}^{(n)}_{\Psi^{(n)},:}\}$ $\leftarrow$ $\mathcal{G}\{\textbf{A}^{(n)}_{\Psi^{(n)},:}\}$
		\STATE $\mathcal{S}\{\textbf{C}^{(n)}_{\Psi^{(n)}}\}$ $\leftarrow$ $\mathcal{R}\{\textbf{C}^{(n)}_{\Psi^{(n)},:}\}$ $\leftarrow$ $\mathcal{R}\{\textbf{A}^{(n)}_{\Psi^{(n)},:}\}$ $\cdot$ $\mathcal{R}\{\textbf{B}^{(n)}\}$
		by Tensor Cores
		\FOR{$k$ from $1$ to $N$ and $k$ $\neq$ $n$}
		\STATE $\mathcal{R}\{\textbf{D}^{(k)}_{\Psi^{(k)},:}\}$ $\leftarrow$ $\mathcal{R}\{\textbf{D}^{(k)}_{\Psi^{(k)},:}\}$ $*$ $\mathcal{S}\{\textbf{C}^{(n)}_{\Psi^{(n)},:}\}$
		\ENDFOR
		\ENDFOR
		\STATE $\mathcal{R}\{\widehat{\textbf{X}}_{\Psi}\}$ $\leftarrow$ $\mathcal{S}\{\textbf{C}^{(1)}_{\Psi^{(1)},:}\}$ $\odot$ $\mathcal{R}\{\textbf{D}^{(1)^{T}}_{\Psi^{(1)},:}\}$ - $\mathcal{G}\{\textbf{X}_{\Psi}\}$
		\FOR{$n$ from $1$ to $N$}
		\STATE $\mathcal{R}\{\textbf{D}^{(n)}_{\Psi^{(n)},:}\}$ $\leftarrow$ $\mathcal{S}\{\textbf{D}^{(n)}_{\Psi^{(n)},:}\}$ $\leftarrow$ $\mathcal{R}\{\textbf{D}^{(n)}_{\Psi^{(n)},:}\}$
		\STATE $\mathcal{R}\{\textbf{E}^{(n)^{T}}_{\Psi^{(n)},:}\}$ $\leftarrow$ $\mathcal{S}\{\textbf{E}^{(n)}_{\Psi^{(n)},:}\}$ $\leftarrow$ 
		$\mathcal{R}\{\widehat{\textbf{X}}_{\Psi}\}$ $\circledast$ 
		$\mathcal{S}\{\textbf{A}^{(n)}_{\Psi^{(n)},:}\}$
		\STATE $\mathcal{S}\{\textbf{E}^{(n)^{T}}_{\Psi^{(n)},:}\textbf{D}^{(n)}_{\Psi^{(n)},:}\}$ $\leftarrow$  $\mathcal{R}\{\textbf{E}^{(n)^{T}}_{\Psi^{(n)},:}\textbf{D}^{(n)}_{\Psi^{(n)},:}\}$ $\leftarrow$ 
		$\mathcal{R}\{\textbf{E}^{(n)^{T}}_{\Psi^{(n)},:}\}$ $\cdot$
		$\mathcal{R}\{\textbf{D}^{(n)}_{\Psi^{(n)},:}\}$ by Tensor Cores
		\STATE $\mathcal{R}\{Grad(\textbf{B}^{(n)})\}$ $\leftarrow$ 
		$\mathcal{R}\{Grad(\textbf{B}^{(n)})\}$ 
		$+$ $\mathcal{S}\{\textbf{E}^{(n)^{T}}_{\Psi^{(n)},:}\textbf{D}^{(n)}_{\Psi^{(n)},:}\}$
		\ENDFOR	
		\ENDFOR
		\FOR{$n$ from $1$ to $N$}
		\STATE $\mathcal{G}\{Grad(\textbf{B}^{(n)})\}$ $\leftarrow$ $\mathcal{G}\{Grad(\textbf{B}^{(n)})\}$ $+$ $\mathcal{R}\{Grad(\textbf{B}^{(n)})\}$
		\ENDFOR	
		\ENDFOR
		\FOR{$n$ from $1$ to $N$}
		\STATE $\mathcal{G}\{\textbf{B}^{(n)}\}$ $\leftarrow$ $\mathcal{G}\{\textbf{B}^{(n)}\}$ 
		$-$ $\gamma_{\textbf{B}}$ $\cdot$
		$\big($ $\mathcal{G}\{Grad(\textbf{B}^{(n)})$ $/$ $|\Omega|$
		$+$$\lambda_{\textbf{B}}$ $\cdot$ $\mathcal{G}\{\textbf{B}^{(n)}\}$ $\big)$		
		\ENDFOR
	\end{algorithmic}
\end{algorithm}

\textbf{Memory Coalescing}:
Global Memory access latency on GPUs can be a significant performance bottleneck, often taking hundreds of clock cycles. To mitigate this issue, GPU architectures employ parallelization techniques to enhance memory access throughput. When a specific location in global memory is accessed, a sequence of consecutive locations that includes the target location is fetched in a single operation. This approach leverages the fact that threads within a warp execute the same instructions simultaneously. In cuFastTuckerPlus, each warp is assigned the responsibility of handling one or more tiles. To optimize memory access patterns, we arrange the factor matrices as $\textbf{A}^{(n)}$, $n$ $\in$ $\{N\}$ and the core matrices as $(\textbf{B}^{(n)})_{p_n,q}$, $p_n$ $\in$ $\{P_n\}$, $q$ $\in$ $\{Q\}$, $n$ $\in$ $\{N\}$ in memory. This arrangement ensures that when a warp reads or writes $\textbf{A}^{(n)}_{\Psi^{(n)},:}$, $n$ $\in$ $\{N\}$ or $\textbf{B}^{(n)}$, $n$ $\in$ $\{N\}$, respectively, it minimizes the number of memory requests required. By reducing memory access overhead, this technique helps enhance the efficiency of cuFastTuckerPlus.

\textbf{Warp Shuffle}:
The Warp Shuffle instruction is a feature that enables direct data exchange between threads within the same warp. It utilizes dedicated hardware, has low latency, and does not require additional memory space. Compared to using shared memory for data exchange, utilizing Warp Shuffle is more efficient. Furthermore, the reduced shared memory usage allows for allocating the saved resources to the on-chip L1 cache, enhancing overall performance.
However, it's important to note that data exchange between different warps within the same block still necessitates the use of shared memory. In scientific computing, the Warp Shuffle instruction is commonly employed for computing operations like the dot product of two vectors or the sum of an array.
In cuFastTuckerPlus, we leverage the Warp Shuffle instruction for matrix multiplication and dot product calculations. For example, we utilize it in expressions such as $\textbf{A}_{\Psi^{(1)},:}^{(1)}\odot(\textbf{B}^{(1)}\textbf{D}_{\Psi^{(1)},:}^{(1)^{T}})$ $\in$ $\mathbb{R}^{M \times 1}$ (line 20 of Algorithm \ref{cuFastTuckerPlus_Factor}), $\textbf{C}_{\Psi^{(1)},:}^{(1)}\odot\textbf{D}_{\Psi^{(1)},:}^{(1)^{T}}$ (line 18 of Algorithm \ref{cuFastTuckerPlus_Core}), $(\textbf{X}_{\Psi}
-\widehat{\textbf{X}}_{\Psi})\circledast(\textbf{D}_{\Psi^{(n)},:}^{(n)}\textbf{B}^{(n)^{T}})$, $n$ $\in$ $\{N\}$ (line 22 of Algorithm \ref{cuFastTuckerPlus_Factor}) and $(\textbf{X}_{\Psi}
-\widehat{\textbf{X}}_{\Psi})\circledast\textbf{A}_{\Psi^{(n)},:}^{(n)}$, $n$ $\in$ $\{N\}$ (line 21 of Algorithm \ref{cuFastTuckerPlus_Core}). These operations involve efficient data exchange within the warp, facilitated by the Warp Shuffle instruction.

\textbf{WMMA API}:
The utilization of the WMMA API enables us to leverage Tensor Cores for performing matrix multiplication computations within the warp. Initially, the matrices are read from global memory or shared memory into the registers of the warp. However, it's important to note that access to the entire matrix can only be made at the warp level, and accessing specific elements within the matrix is not possible at the thread level. Once the Tensor Cores complete their calculations, the results stored in registers can be written back to global memory or shared memory.
In cuFastTuckerPlus, we leverage Tensor Cores to compute the following matrix multiplications: $\textbf{A}_{\Psi^{(n)},:}^{(n)}$ $\cdot$ $\textbf{B}^{(n)}$, $n$ $\in$ $\{N\}$ (line 11 of Algorithm \ref{cuFastTuckerPlus_Factor} and line 13 of Algorithm \ref{cuFastTuckerPlus_Core}), 
$\textbf{B}^{(n)}$ $\cdot$  $\textbf{D}_{\Psi^{(n)},:}^{(n)^{T}}$ , $n$ $\in$ $\{N\}$  (line 18 of Algorithm \ref{cuFastTuckerPlus_Factor}) and
$\textbf{E}_{\Psi^{(n)},:}^{(n)^{T}}$ $\cdot$ $\textbf{B}^{(n)}$, $n$ $\in$ $\{N\}$ (line 22 of Algorithm \ref{cuFastTuckerPlus_Core}). These matrix multiplication operations are efficiently performed using the Tensor Cores. However, it's worth mentioning that these operations can also be carried out using Warp Shuffle with CUDA Cores. When employing CUDA Cores for tile multiplication, two workers are responsible for calculating the upper and lower parts of the tile, and each worker computes a dot product at a time.

\textbf{Loop Unrolling}:
Loop unrolling is an instruction-level optimization technique that enhances code performance by sacrificing programming complexity. By unrolling loops in CUDA, we can reduce instruction costs and introduce more independently dispatched instructions. This approach increases the number of concurrent operations in the pipeline, leading to improved instruction and memory bandwidth.
In our implementation, we utilize the $\#pragma$ $unroll$ directive to perform loop unrolling, which doesn't significantly increase the program's complexity. Additionally, it's important to note that CUDA does not support register arrays, but fixed-length short arrays can be stored in registers. When an array is too long or has an indeterminate index, the compiler may store it in global memory. However, by unrolling the loop, we can eliminate the indeterminate index of the array, allowing the compiler to allocate the fixed-length short array into registers. This optimization technique enables faster access to arrays stored in registers rather than relying on shared memory for certain computational tasks. In cuFastTuckerPlus, we leverage this trick to store specific parameters in registers, resulting in improved performance.

\textbf{Shared Memory}:
Shared memory is a high-bandwidth and low-latency on-chip memory located on the SM. It is allocated in blocks and shared by all threads within a block, persisting throughout the lifetime of the block. Shared memory serves multiple purposes, including reducing access to global memory by storing frequently accessed parameters (hotspot parameters) and facilitating communication among threads in different warps within the same block.
In cuFastTuckerPlus, we allocate exclusive shared memory for each warp in the block. This shared memory is used to store various parameters, such as $\textbf{A}_{\Psi^{(n)},:}^{(n)}$, $n$ $\in$ $\{N\}$ (line 10 of Algorithm \ref{cuFastTuckerPlus_Factor} and line 12 of Algorithm \ref{cuFastTuckerPlus_Core}), 
$\textbf{C}_{\Psi^{(n)},:}^{(n)}$, $n$ $\in$ $\{N\}$ (line 11 of Algorithm \ref{cuFastTuckerPlus_Factor} and line 13 of Algorithm \ref{cuFastTuckerPlus_Core}), 
$\textbf{D}_{\Psi^{(n)},:}^{(n)}$, $n$ $\in$ $\{N\}$ (line 17 of Algorithm \ref{cuFastTuckerPlus_Factor} and line 20 of Algorithm \ref{cuFastTuckerPlus_Core}), 
$\textbf{B}^{(n)}\textbf{D}_{\Psi^{(n)},:}^{(n)^{T}}$, $n$ $\in$ $\{N\}$ (line 18 of Algorithm \ref{cuFastTuckerPlus_Factor}) and
$\textbf{E}^{(n)^{T}}_{\Psi^{(n)},:}\textbf{D}^{(n)}_{\Psi^{(n)},:}$, $n$ $\in$ $\{N\}$ (line 22 of Algorithm \ref{cuFastTuckerPlus_Core}). These parameters are utilized as inputs and outputs of Tensor Cores. As these parameters are shared across different time steps, they do not consume a significant amount of shared memory. By leveraging shared memory efficiently, we can reduce global memory access and enhance the overall performance of cuFastTuckerPlus.

\textbf{Register}:
Registers are the fastest memory in the GPU, offering quick access to data for calculations, reading, and writing. However, registers are limited in size and are a valuable resource. It is crucial to allocate parameters efficiently to registers to maximize performance while ensuring enough warps survive in the Streaming Multiprocessor (SM) to maintain high data throughput.
In cuFastTuckerPlus, we prioritize frequently used parameters and store them in registers without occupying excessive register space. Intermediate variables and the results of computing $\textbf{D}_{\Psi^{(n)},:}^{(n)}$, $n$ $\in$ $\{N\}$ (lines 7, 13 of Algorithm \ref{cuFastTuckerPlus_Factor} and lines 9, 15 of Algorithm \ref{cuFastTuckerPlus_Core}) are stored in registers. We also use registers to store and accumulate the gradient of $\textbf{B}^{(n)}$, $n$ $\in$ $\{N\}$ (line 8 of Algorithm \ref{cuFastTuckerPlus_Core}), which is later written to global memory to minimize global memory writes.
Furthermore, we utilize registers with Tensor Cores to store $\textbf{A}_{\Psi^{(n)},:}^{(n)}$, $n$ $\in$ $\{N\}$ (line 10 of Algorithm \ref{cuFastTuckerPlus_Factor} and line 12 of Algorithm \ref{cuFastTuckerPlus_Core}), $\textbf{B}^{(n)}$, $n$ $\in$ $\{N\}$ (line 2 of Algorithm \ref{cuFastTuckerPlus_Factor} and line 2 of Algorithm \ref{cuFastTuckerPlus_Core}) and $\textbf{D}_{\Psi^{(n)},:}^{(n)}$, $n$ $\in$ $\{N\}$. Correspondingly, the calculated $\textbf{C}_{\Psi^{(n)},:}^{(n)}$, $n$ $\in$ $\{N\}$ (line 11 of Algorithm \ref{cuFastTuckerPlus_Factor} and line 13 of Algorithm \ref{cuFastTuckerPlus_Core}), 
$\textbf{B}^{(n)}\textbf{D}^{(n)^{T}}_{\Psi^{(n)},:}$, $n$ $\in$ $\{N\}$ (line 18 of Algorithm \ref{cuFastTuckerPlus_Factor})
and $\textbf{E}^{(n)^{T}}_{\Psi^{(n)},:}\textbf{D}^{(n)}_{\Psi^{(n)},:}$, $n$ $\in$ $\{N\}$ (line 22 of Algorithm \ref{cuFastTuckerPlus_Core}) are also stored in registers with Tensor Cores. Additionally, other intermediate processes within the calculation process are performed in registers to take advantage of their speed and minimize memory access.
By strategically utilizing registers for storing parameters and intermediate values, cuFastTuckerPlus optimizes performance and reduces the reliance on shared memory and global memory, resulting in efficient computations.

\textbf{Read-only Data Cache}:
In cuFastTuckerPlus, we take advantage of the fact that the values of $\textbf{A}_{\Psi^{(n)},:}^{(n)}$, $n$ $\in$ $\{N\}$ remain unchanged between the warp reading and updating stages. To optimize memory access and reduce shared memory usage, we store $\textbf{A}_{\Psi^{(n)},:}^{(n)}$, $n$ $\in$ $\{N\}$ (line 10 of Algorithm \ref{cuFastTuckerPlus_Factor} and line 12 of Algorithm \ref{cuFastTuckerPlus_Core}) in the read-only cache.
The read-only cache, also known as the constant cache, is a specialized cache in NVIDIA GPUs designed for storing read-only data. By storing $\textbf{A}_{\Psi^{(n)},:}^{(n)}$, $n$ $\in$ $\{N\}$ in the read-only cache, we enable faster memory access for the warp reading stage.
To optimize read-only memory access, we utilize the $\_\_ldg$ modifier in NVIDIA GPUs. The $\_\_ldg$ modifier is a CUDA compiler intrinsic that hints the compiler to use the read-only cache for memory loads. It allows for more efficient and optimized access to read-only memory, improving performance in scenarios where the data remains unchanged.
By leveraging the read-only cache and utilizing the $\_\_ldg$ modifier, cuFastTuckerPlus enhances memory access efficiency when reading $\textbf{A}_{\Psi^{(n)},:}^{(n)}$, $n$ $\in$ $\{N\}$ contributing to overall performance optimization.

\textbf{Atomic Operation}:
In cuFastTuckerPlus, when updating the core matrices, each warp independently accumulates the gradient of $\textbf{B}^{(n)}$, $n$ $\in$ $\{N\}$ using registers. However, when it comes to writing the accumulated gradients to global memory, we need to ensure data consistency among multiple threads to avoid conflicts.
To address this, cuFastTuckerPlus utilizes the $atomicAdd$ function (line 27 of Algorithm \ref{cuFastTuckerPlus_Core}) to perform atomic addition, ensuring that the updates to the global memory are done in a synchronized and consistent manner.
The $atomicAdd$ function is an atomic operation provided by CUDA that performs an atomic addition of a value to the specified memory location. It guarantees that the addition operation is executed atomically without any interference from other threads. This prevents data races and ensures that the updates to the global memory are correctly accumulated.
By using atomicAdd, cuFastTuckerPlus ensures the integrity and consistency of the gradient updates performed by different warps when writing the gradients of $\textbf{B}^{(n)}$, $n$ $\in$ $\{N\}$ to global memory, avoiding conflicts and data corruption.

\section{Experiments} \label{Section_Experiments}

We present experimental results to answer the following questions.

\begin{enumerate}
	
	\item \emph{Convergence Speed}: Does the non-convex optimization SGD algorithm cuFastTuckerPlus demonstrate faster convergence compared to convex optimization SGD algorithms?
	
	\item \emph{Single Iteration Running Time}: Does a single iteration of cuFastTuckerPlus exhibit reduced execution time in comparison to other algorithms?
	
	\item \emph{Memory Access Overhead}: What advantages does cuFastTuckerPlus offer in terms of memory access overhead?
	
	\item \emph{Tensor Cores' Impact}: How does cuFastTuckerPlus's acceleration performance fare when utilizing Tensor Cores? Additionally, what is the acceleration performance of other algorithms when employing Tensor Cores?
	
	\item \emph{Calculation or Storage}: Which approach yields superior performance: precomputing and storing the intermediate matrices $\textbf{C}^{(n)}$, $n$ $\in$ $\{N\}$ in advance, or computing $\textbf{C}_{\Psi^{(n)},:}^{(n)}$, $n$ $\in$ $\{N\}$ temporarily when needed?
	
	\item \emph{Running Time in Various Parameters}: What is the impact of different parameters on the running time of cuFastTuckerPlus?
	
\end{enumerate}

We provide a detailed description of the datasets and experimental settings in Section \ref{Section_Experimental_Settings}. The questions posed in this study are addressed as follows: the speed of convergence is investigated in Section \ref{Section_Convergence_Speed}, the single iteration running time of the algorithms is analyzed in Section \ref{Section_Single_iteration_running_time_of_the_algorithm}, the memory access overhead is evaluated in Section \ref{Section_Memory_Access_Overhead}, the effect of the Tensor Cores on cuFastTuckerPlus and other algorithms is examined in Section \ref{Section_The_Speedup_of_Algorithms by_Tensor_Cores}, the choice between calculation and storage approaches is discussed in Section \ref{Section_Replace_memory_access_with_calculation}, and the running time of cuFastTuckerPlus under various parameter settings is explored in Section \ref{Section_The_Running_Time_of_cuFastTuckerPlus_in_Various_Parameters}.

\subsection{Experimental Settings} \label{Section_Experimental_Settings}

\begin{table}
	\caption{Datasets}
	\tiny
	\centering
	\subtable[Real World Datasets]{
		\centering
		\begin{tabular}{c|cc}
			\hline
			\hline
			& Netflix        & Yahoo!Music       \\
			\hline
			$I_1$           & 480, 189       & 1, 000, 990       \\
			$I_2$           & 17, 770        & 624, 961          \\
			$I_3$           & 2, 182         & 3, 075            \\
			$|\Omega|$      & 99, 072, 112   & 250, 272, 286     \\
			$|\Gamma|$      & 1, 408, 395    & 2, 527, 989       \\
			Max Value       & 5              & 5                 \\
			Min Value       & 1              & 0.025             \\
			\hline
			\hline
		\end{tabular}
		\label{data_sets_real_world}
	}
	\qquad
	\subtable[Synthetic Datasets]{  
		\centering      
		\begin{tabular}{c|c}
			\hline
			\hline
			& Synthetic          \\
			\hline
			order         	& 3,4,5,6,7,8,9,10   \\
			$I$             & 10, 000            \\
			$|\Omega|$      & 100,000,000        \\
			Max Value       & 5                  \\
			Min Value       & 1                  \\
			\hline
			\hline
		\end{tabular}
		\label{data_sets_synthetic}
	}
\end{table}

\begin{enumerate}
	
	\item \emph{Datasets}: 	
	We utilize a combination of real-world datasets and synthetic datasets to evaluate the convergence and performance of the algorithms. For the real-world datasets, we employ two well-known rating datasets: Netflix\footnotemark[1] \footnotetext[1]{https://www.netflixprize.com/} 
	and Yahoo!music \footnotemark[2] \footnotetext[2]{https://webscope.sandbox.yahoo.com/}, commonly used in the recommendation system domain. The Netflix dataset is a $3$-order tensor, where the dimensions represent users, movies, and time, respectively. The values within the tensor represent movie ratings provided by users at specific times. Similarly, the Yahoo!Music dataset is also a $3$-order tensor, with dimensions representing users, music, and time, and values representing music ratings provided by users at specific times. Both datasets are divided into a trainset, denoted as $\Omega$, and a testset, denoted as $\Gamma$. We train the parameters on the trainset $\Omega$ and evaluate their performance on the testset $\Gamma$.
	In addition to the real-world datasets, we construct synthetic datasets consisting of $8$ sparse tensors with orders ranging from $3$ to $10$. Each tensor contains a total of $100,000,000$ elements, with each dimension having a size of $10,000$. These synthetic tensors allow us to assess the performance of the algorithms on HHLST.
	For further details on the characteristics of the real-world and synthetic datasets, please refer to Tables \ref{data_sets_real_world} and \ref{data_sets_synthetic}, respectively.
	
	\item \emph{Contrasting algorithms}: 
	It has been demonstrated that Sparse Tucker Decomposition algorithms, such as P-Tucker, Vest, SGD\_\_Tucker, ParTi, GTA, and cuTucker, are not suitable for handling HHLST. In this study, we compare our proposed cuFastTuckerPlus algorithm with the SOTA Sparse FastTucker Decomposition algorithms, namely cuFastTucker and cuFasterTucker. Below are brief descriptions of each method:
	
	\begin{itemize}	
		
		\item \textbf{cuFastTucker}: A large-scale parallel Sparse FastTucker Decomposition algorithm based on Algorithm \ref{Algorithm_FastTucker} implemented on the CUDA platform.
		
		\item \textbf{cuFasterTucker}: A large-scale parallel Sparse FastTucker Decomposition algorithm based on Algorithm \ref{Algorithm_FasterTucker} implemented on the CUDA platform.
		
		\item \textbf{cuFasterTuckerCOO}: A variant of the cuFasterTucker algorithm that utilizes the COO tensor storage format. Consequently, it does not reduce the computation of shared invariant intermediate variables.
		
		\item \textbf{cuFastTuckerPlus\_CC}: A basic implementation of our proposed Algorithm \ref{Algorithm_FastTuckerPlus} on the CUDA platform using CUDA Cores, without acceleration by Tensor Cores.
		
		\item \textbf{cuFastTucker\_TC}: A variant of cuFastTucker that leverages Tensor Cores for acceleration.
		
		\item \textbf{cuFasterTucker\_TC}: A variant of cuFasterTucker that takes advantage of Tensor Cores for acceleration.
		
		\item \textbf{cuFasterTuckerCOO\_TC}: A variant of cuFasterTuckerCOO that utilizes Tensor Cores for acceleration.
		
		\item \textbf{cuFastTuckerPlus}: Our efficient non-convex optimization algorithm proposed in this work, which utilizes Tensor Cores for acceleration on the CUDA platform.
		
	\end{itemize}	
	
	\emph{Environment:} The implementation of cuFastTuckerPlus is done in C/C++ using CUDA. The experiments for cuFastTucker, cuFasterTucker, cuFastTuckerPlus, and their respective variants are conducted on an \textbf{NVIDIA GeForce RTX 3080Ti GPU} equipped with $10,240$ CUDA Cores, $320$ Tensor Cores, and $12GB$ of graphic memory. All algorithms utilize optimal parameters and are executed without interference from other tasks. For the sake of consistency, we set $J_n=16$, $n$ $\in$ $\{N\}$ and $R=16$ across all algorithms. Furthermore, each algorithm is iterated for a total of $50$ times.
	
\end{enumerate}

\subsection{Convergence Speed} \label{Section_Convergence_Speed}

We utilize the Root Mean Square Error (RMSE) and Mean Absolute Error (MAE) of the testset $\Gamma$ to assess the accuracy of the algorithms. To ensure fairness, all algorithms are initialized with the same random starting point, and the RMSE and MAE are recorded after each iteration. Figure \ref{convergence} illustrates the convergence curves of cuFastTuckerPlus and other algorithms on the Netflix and Yahoo!Music datasets.
For the Netflix dataset, the baseline test RMSE and test MAE are $0.95$ and $0.75$ respectively. As for the Yahoo!Music dataset, the baseline test RMSE and test MAE are $1.20$ and $0.90$ respectively. All algorithms demonstrate the capability to converge to the baseline values.
Notably, cuFastTuckerPlus and cuFastTuckerPlus\_CC exhibit faster convergence compared to the other algorithms, while also achieving greater accuracy. These results indicate that Sparse FastTucker decomposition benefits from a simple convergence landscapes, and the non-convex optimization-based SGD approach leads to faster convergence.

\begin{figure*}[htbp]
	\centering
	\subfigure[RMSE on Netflix]{
		\label{convergence(a)}
		\includegraphics[width=1.5in]{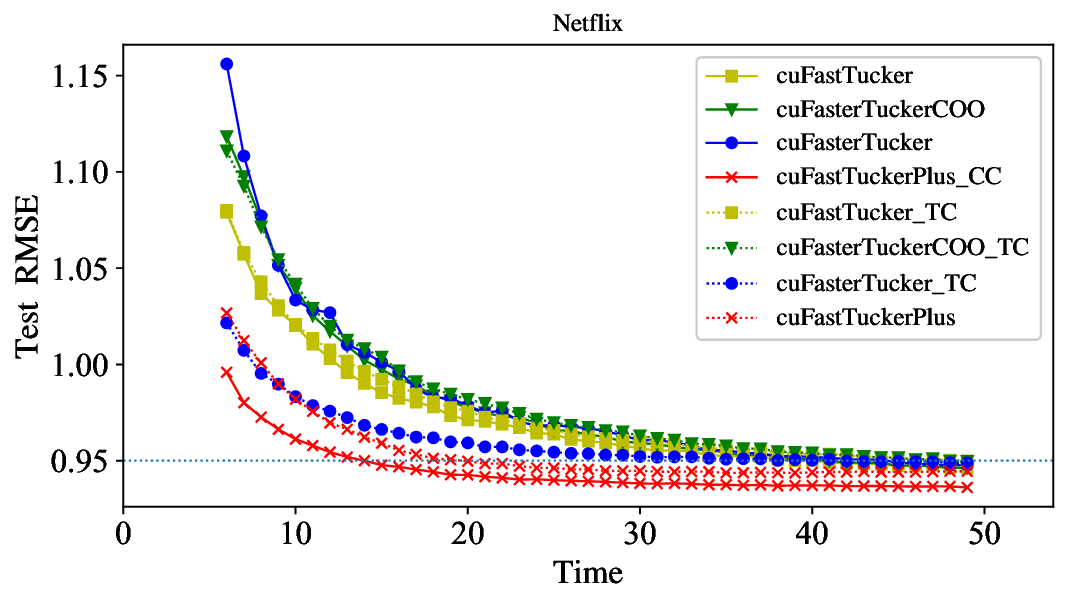}}
	~
	\subfigure[MAE on Netflix]{
		\label{convergence(b)}
		\includegraphics[width=1.5in]{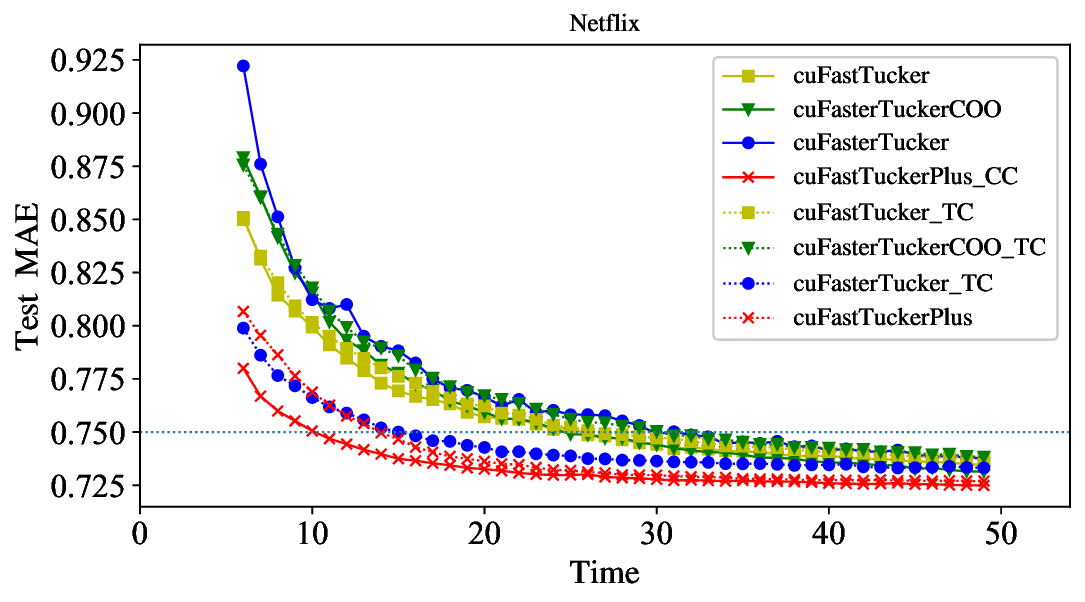}}
	~
	\subfigure[RMSE on Yahoo!Music]{
		\label{convergence(c)}
		\includegraphics[width=1.5in]{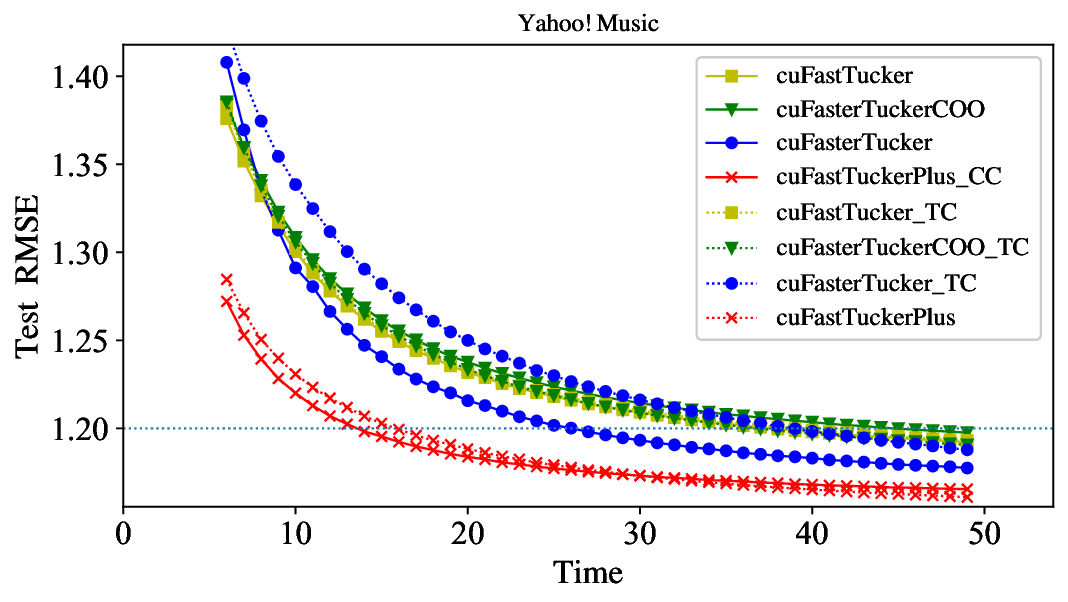}}
	~
	\subfigure[MAE on Yahoo!Music]{
		\label{convergence(d)}
		\includegraphics[width=1.5in]{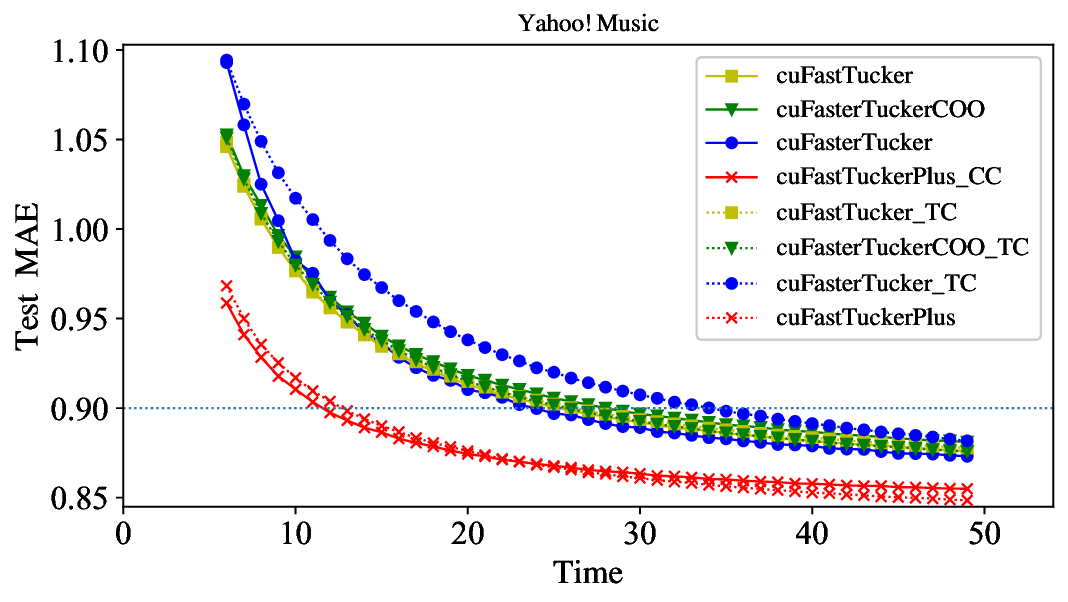}}
	\caption{The convergence curves of cuFastTuckerPlus and other algorithms on Netflix and Yahoo!Music datasets.}
	\label{convergence}
\end{figure*}

\subsection{Single iteration running time of the algorithm} \label{Section_Single_iteration_running_time_of_the_algorithm}

To evaluate the efficiency of the algorithms, we utilize the single iteration running time as a metric. Table \ref{time} provides details on the running time for a single iteration of cuFastTuckerPlus and other algorithms on the Netflix and Yahoo!Music datasets.
It is observed that cuFastTuckerPlus\_CC exhibits a longer running time compared to cuFasterTucker and cuFasterTuckerCOO. However, it still achieves a significant speedup of approximately $3.0X$ compared to cuFastTucker. Additionally, cuFastTuckerPlus exhibits the lowest running time among all the algorithms.
In terms of the synthetic datasets, as depicted in Figure \ref{synthetic_time}, cuFastTuckerPlus\_CC lags behind cuFasterTucker and cuFasterTuckerCOO in terms of running time. However, cuFastTuckerPlus showcases significantly lower running times compared to all other algorithms.

\begin{table}
	\caption{The running time (in seconds) of a single iteration for cuFastTuckerPlus and other algorithms on Netflix and Yahoo!Music datasets, as well as the speedup comparison with the baseline cuFastTucker.}
	\tiny
	\centering
	\label{time}
	\subtable[The process of updating the factor matrices]{
		\centering
		\begin{tabular}{c|ccc}
			\hline \hline
			Algorithm               & Netflix            & Yahoo!Music           \\
			\hline
			cuFastTucker            & 1.084079           & 2.719274              \\
			cuFasterTucker          & 0.123575 (8.77X)   & 0.300970 (9.04X)      \\
			cuFasterTuckerCOO       & 0.133995 (8.09X)   & 0.349147 (7.79X)      \\
			cuFastTuckerPlus        & 0.339884 (3.19X)   & 0.839495 (3.24X)      \\
			cuFastTucker\_TC        & 0.107342 (10.10X)  & 0.314898 (8.64X)      \\
			cuFasterTucker\_TC      & 0.127334 (8.51X)   & 0.344562 (7.89X)      \\
			cuFasterTuckerCOO\_TC   & 0.091440 (11.86X)  & 0.281549 (9.66X)      \\
			cuFastTuckerPlus\_TC    & 0.044434 (24.40X)  & 0.146161 (18.60X)     \\
			\hline \hline
		\end{tabular}
		\label{time_factor}
	}
	\qquad
	\subtable[The process of updating the core matrices]{  
		\centering      
		\begin{tabular}{c|ccc}
			\hline \hline
			Algorithm               & Netflix            & Yahoo!Music           \\
			\hline
			cuFastTucker            & 1.290737           & 3.259773              \\
			cuFasterTucker          & 0.136375 (9.46X)   & 0.332282 (9.81X)      \\
			cuFasterTuckerCOO       & 0.222538 (5.80X)   & 0.563203 (5.79X)      \\
			cuFasterTuckerPlus      & 0.443042 (2.91X)   & 1.118276 (2.91X)      \\
			cuFastTucker\_TC        & 0.106555 (12.11X)  & 0.292456 (11.15X)     \\
			cuFasterTucker\_TC      & 0.178263 (7.24X)   & 0.470346 (6.93X)      \\
			cuFasterTuckerCOO\_TC   & 0.102614 (12.58X)  & 0.304671 (10.70X)     \\
			cuFasterTuckerPlus\_TC  & 0.030067 (42.93X)  & 0.072919 (44.70X)     \\
			\hline \hline
		\end{tabular}
		\label{time_core}
	}
\end{table}

\begin{figure}[htbp]
	\centering
	\subfigure[The process of updating the factor matrices]{
		\label{synthetic_time_factor}
		\includegraphics[width=1.5in]{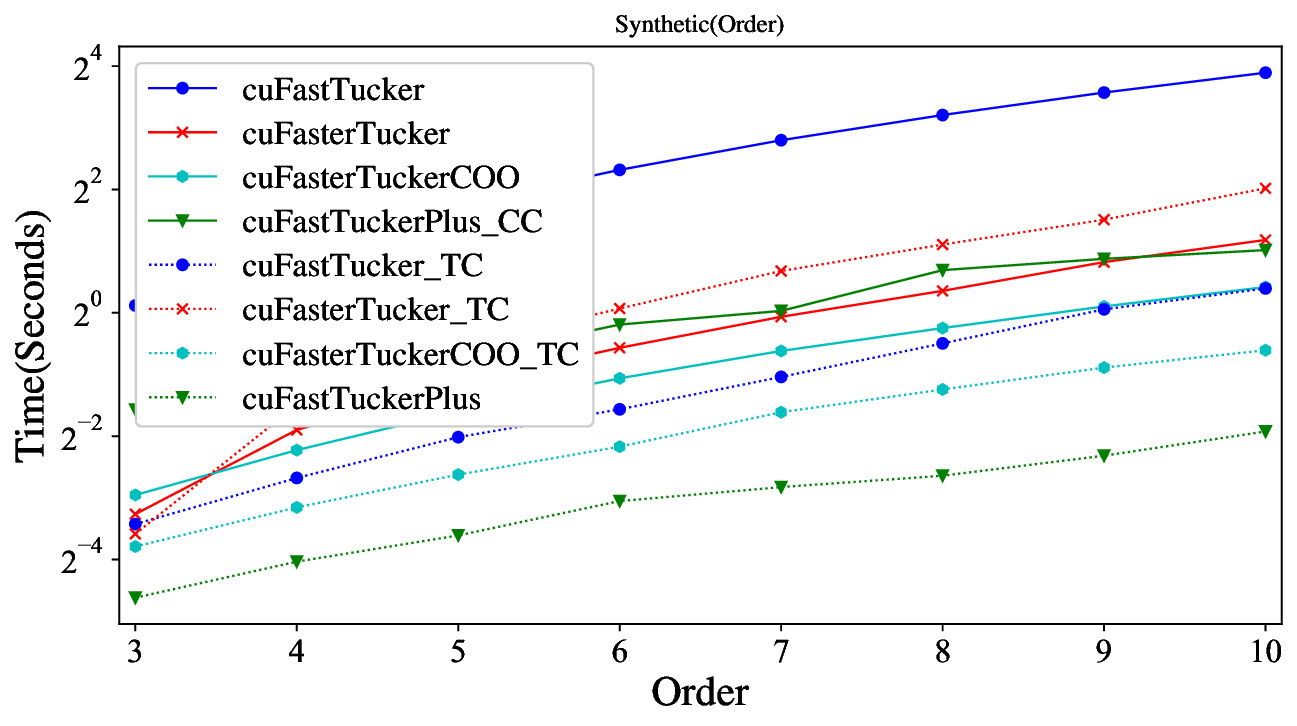}}
	~
	\subfigure[The process of updating the core matrices]{
		\label{synthetic_time_core}
		\includegraphics[width=1.5in]{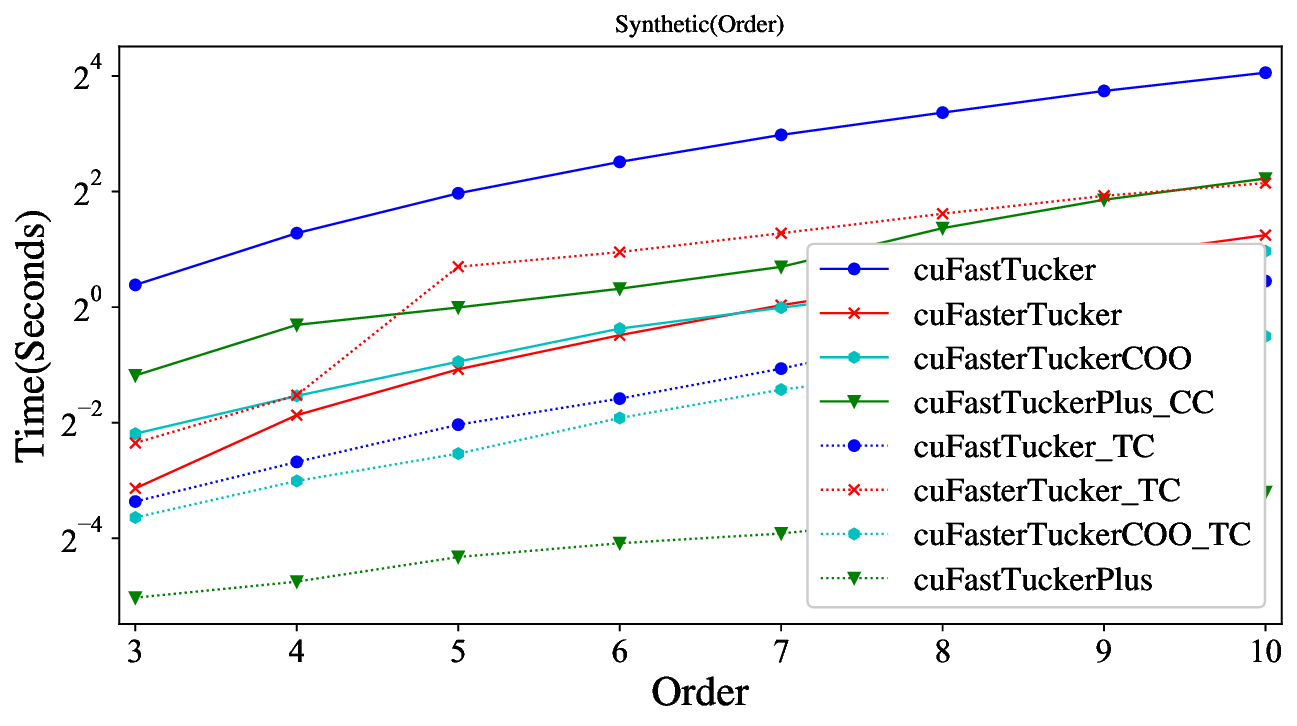}}
	\caption{The running time (in seconds) of cuFastTuckerPlus and other algorithms on synthesis datasets.}
	\label{synthetic_time}
\end{figure}

\subsection{Memory Access Overhead} \label{Section_Memory_Access_Overhead}

Table \ref{real_world_dataset_memory} presents the time required for cuFastTuckerPlus and other algorithms to read parameters from global memory on real-world datasets. Furthermore, Figure \ref{synthetic_dataset_memory} illustrates the time taken by cuFastTuckerPlus and other algorithms to read parameters from global memory on synthetic datasets. The observations from both the table and the figure demonstrate that cuFastTucker exhibits the longest memory access time. For sparse tensors of $3$-order, cuFasterTucker requires less memory access time compared to cuFasterTuckerCOO, while for sparse tensors of $4$-order and above, it takes longer. This discrepancy can be attributed to the higher sparsity of higher-order sparse tensors in the synthetic dataset. Notably, cuFastTuckerPlus showcases the shortest memory access time. Furthermore, as the order of the sparse tensor increases, the memory access time of cuFastTuckerPlus exhibits the slowest growth rate.

\begin{table}
	\caption{The memory access time (in seconds) for cuFastTuckerPlus and other algorithms on Netflix and Yahoo!Music datasets.}
	\label{real_world_dataset_memory}
	\centering
	\subtable[The process of updating the factor matrices]{
		\centering
		\begin{tabular}{c|ccc}
			\hline \hline
			Algorithm               & Netflix    & Yahoo!Music      \\
			\hline
			cuFastTucker            & 0.517661   & 1.326015         \\
			cuFasterTucker          & 0.050513   & 0.130205         \\
			cuFasterTuckerCOO       & 0.063294   & 0.213729         \\
			cuFastTuckerPlus        & 0.022015   & 0.069812         \\
			\hline \hline
		\end{tabular}
		\label{real_world_dataset_memory_factor}
	}
	\qquad
	\subtable[The process of updating the core matrices]{  
		\centering      
		\begin{tabular}{c|ccc}
			\hline \hline
			Algorithm               & Netflix    & Yahoo!Music      \\
			\hline
			cuFastTucker            & 0.584005   & 1.466973         \\
			cuFasterTucker          & 0.050248   & 0.130966         \\
			cuFasterTuckerCOO       & 0.083033   & 0.280937         \\
			cuFasterTuckerPlus      & 0.020940   & 0.069255         \\
			\hline \hline
		\end{tabular}
		\label{real_world_dataset_memory_core}
	}
\end{table}

\begin{figure}[htbp]
	\centering
	\subfigure[The process of updating the factor matrices]{
		\label{synthetic_dataset_memory_factor}
		\includegraphics[width=1.5in]{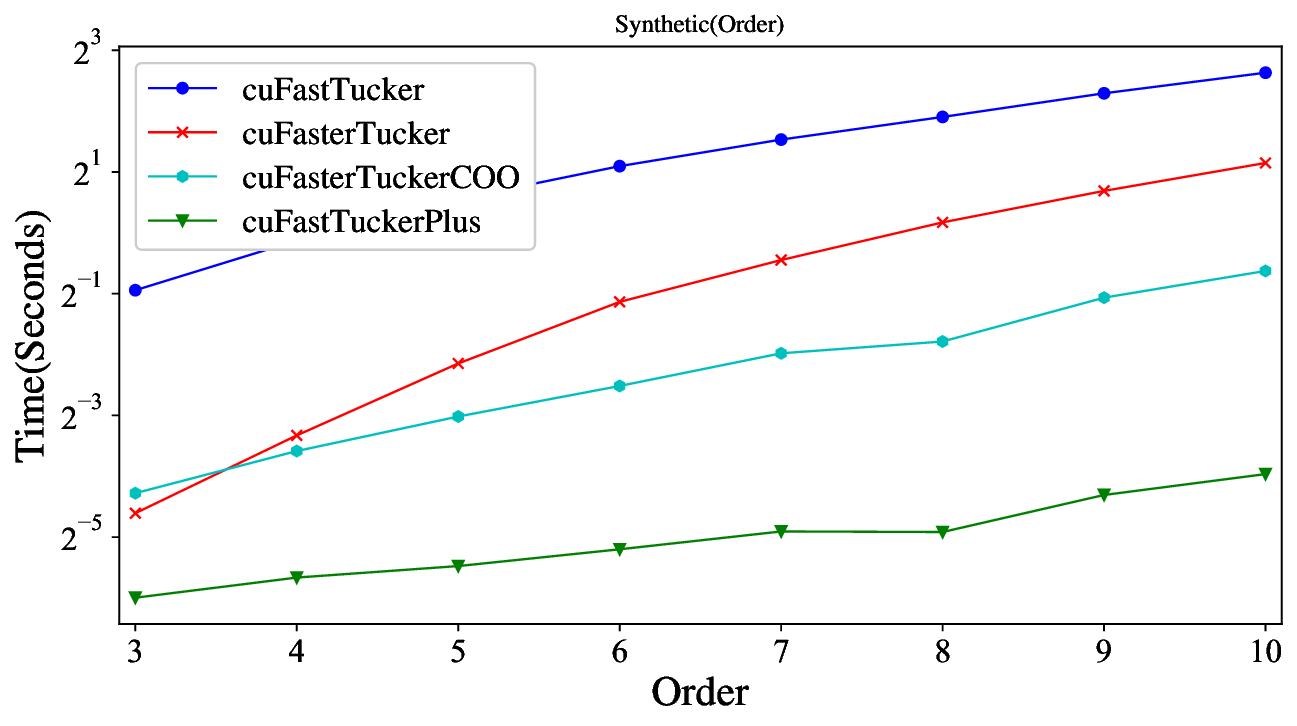}}
	~
	\subfigure[The process of updating the core matrices]{
		\label{synthetic_dataset_memory_core}
		\includegraphics[width=1.5in]{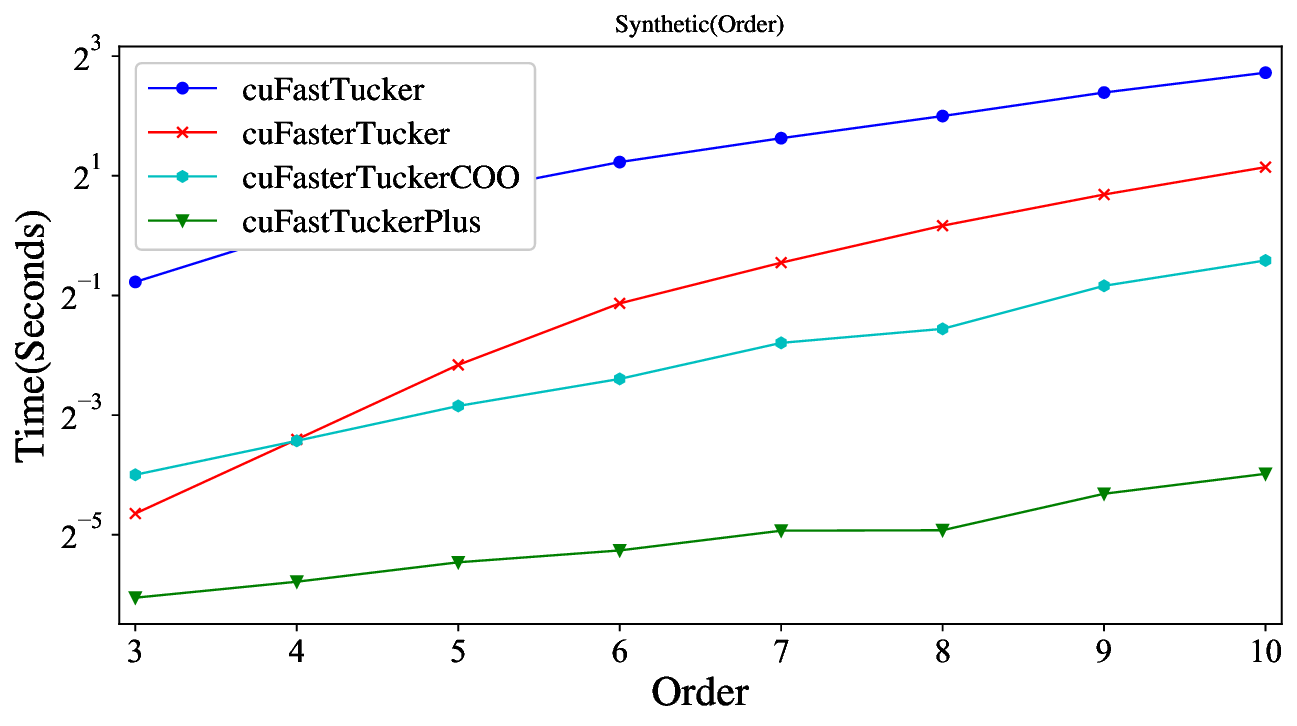}}
	\caption{The memory access time (in seconds) for cuFastTuckerPlus and other algorithms on synthesis datasets.}
	\label{synthetic_dataset_memory}
\end{figure}

\subsection{The Speedup of Algorithms by Tensor Cores} \label{Section_The_Speedup_of_Algorithms by_Tensor_Cores}

Table \ref{real_world_dataset_speedup} provides the speedup achieved by cuFastTuckerPlus and other algorithms after employing Tensor Cores on real-world datasets. Additionally, Figure \ref{synthetic_dataset_speedup} illustrates the speedup attained by cuFastTuckerPlus and other algorithms with Tensor Cores on synthetic datasets. The speedup is calculated as the ratio of the single iteration time required by the algorithm using CUDA Cores to the single iteration time required by the algorithm accelerated by Tensor Cores. Analysis of Table \ref{real_world_dataset_speedup} and Figure \ref{synthetic_dataset_speedup} reveals significant acceleration for cuFastTucker\_TC and cuFastTuckerPlus. Conversely, cuFasterTucker\_TC experiences an increase in runtime, while cuFasterTuckerCOO\_TC achieves a slight speedup. Notably, when updating the factor matrices, cuFastTucker\_TC demonstrates higher speedup compared to cuFastTuckerPlus. This discrepancy arises due to the higher computational complexity of cuFastTucker, with a larger portion of matrix calculations eligible for acceleration by Tensor Cores. On the other hand, when updating the core matrices, cuFastTuckerPlus outperforms cuFastTucker\_TC in terms of speedup. This distinction is attributed to cuFastTuckerPlus\_CC storing more parameters than cuFastTucker, leading to performance degradation in cuFastTuckerPlus and thereby making the acceleration more pronounced. Both cuFasterTucker and cuFasterTuckerCOO primarily rely on reading $\textbf{c}^{(n)}_{i_n,:}, n \in\{N\}$ from global memory rather than utilizing Tensor Cores for calculations. Consequently, the matrix calculations suitable for Tensor Cores acceleration are minimal. Moreover, cuFasterTucker\_TC experiences imbalanced performance due to the use of Tensor Cores, resulting in slower performance compared to cuFasterTuckerCOO\_TC.

\begin{table}
	\caption{The speedup achieved by cuFastTuckerPlus and other algorithms when utilizing Tensor Cores on the Netflix and Yahoo!Music datasets.}
	\centering
	\subtable[The process of updating the factor matrices]{
		\centering
		\begin{tabular}{c|ccc}
			\hline \hline
			Algorithm               & Netflix    & Yahoo!Music   \\
			\hline
			cuFastTucker\_TC        & 10.10X     & 8.64X         \\
			cuFasterTucker\_TC      & 0.97X      & 0.87X         \\
			cuFasterTuckerCOO\_TC   & 1.47X      & 1.24X         \\
			cuFastTuckerPlus        & 7.64X      & 5.74X         \\
			\hline \hline
		\end{tabular}
		\label{real_world_dataset_speedup_factor}
	}
	\qquad
	\subtable[The process of updating the core matrices]{  
		\centering      
		\begin{tabular}{c|ccc}
			\hline \hline
			Algorithm               & Netflix    & Yahoo!Music   \\
			\hline
			cuFastTucker\_TC        & 12.11X     & 11.15X        \\
			cuFasterTucker\_TC      & 0.77X      & 0.71X         \\
			cuFasterTuckerCOO\_TC   & 2.17X      & 1.85X         \\
			cuFasterTuckerPlus      & 14.74X     & 15.34X        \\
			\hline \hline
		\end{tabular}
		\label{real_world_dataset_speedup_core}		
	}
	\label{real_world_dataset_speedup}
\end{table}

\begin{figure}[htbp]
	\centering
	\subfigure[The process of updating the factor matrices]{
		\label{synthetic_dataset_speedup_factor}
		\includegraphics[width=1.5in]{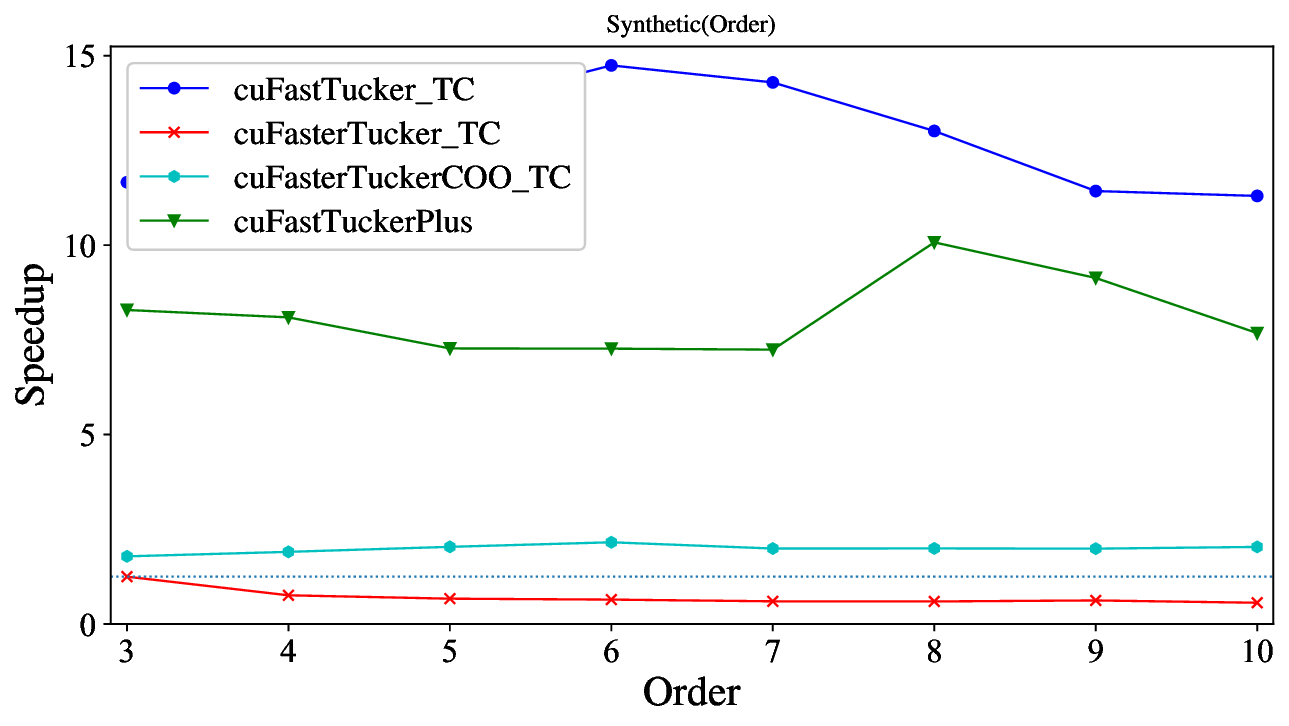}}
	~
	\subfigure[The process of updating the core matrices]{
		\label{synthetic_dataset_speedup_core}
		\includegraphics[width=1.5in]{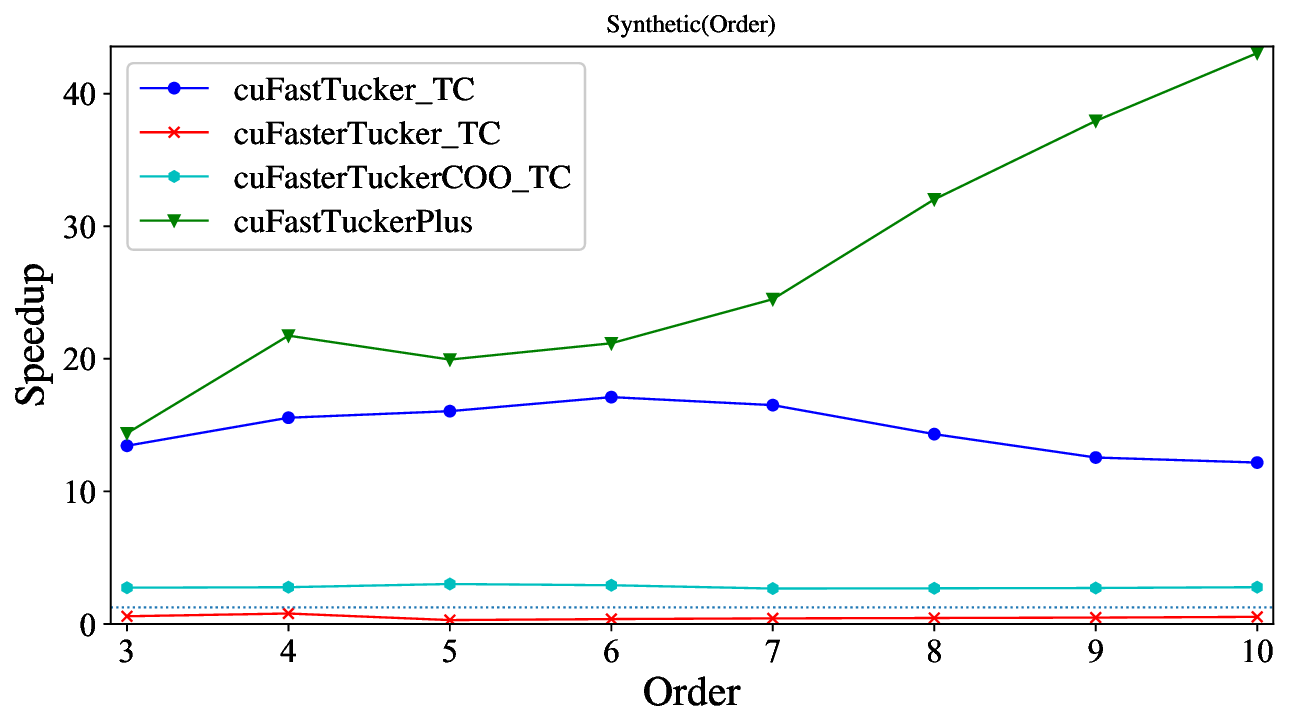}}
	\caption{The speedup achieved by cuFastTuckerPlus and other algorithms when utilizing Tensor Cores on the synthesis datasets.}
	\label{synthetic_dataset_speedup}
\end{figure}

\subsection{Replace memory access with calculation} \label{Section_Replace_memory_access_with_calculation}

When updating the factor matrices in cuFastTuckerPlus and cuFastTuckerPlusTC, the factor matrices $\textbf{A}^{(n)},$ $n \in \{N\}$ are updated dynamically, which means that $\textbf{c}^{(n)}_{\Psi^{(n)},:}, n \in\{N\}$ can only be computed in real-time during the update process.
However, when updating the core matrices, since $\textbf{B}^{(n)},$ $n \in \{N\}$ is updated at the end after accumulating gradients, it is possible to pre-compute $\textbf{c}^{(n)}_{\Psi^{(n)},:}, n \in\{N\}$ and store it in memory. Then, during the gradient accumulation process, $\textbf{C}^{(n)}_{\Psi^{(n)},:}, n \in\{N\}$ can be directly read from memory, avoiding redundant computations.
To explore this further, we consider two schemes in cuFastTuckerPlus and cuFastTuckerPlusTC: one scheme involves pre-computing $\textbf{C}^{(n)}, n \in\{N\}$ and then reading $\textbf{C}^{(n)}_{\Psi^{(n)},:}, n \in\{N\}$, while the other scheme involves real-time computation of $\textbf{C}^{(n)}_{\Psi^{(n)},:}, n \in\{N\}$. We denote these schemes as $cuFastTuckerPlus\_CC$ $(Storage)$ and $cuFastTuckerPlus\_CC$ $(Calculation)$ for cuFastTuckerPlus\_CC, and $cuFastTuckerPlus$ $(Storage)$ and $cuFastTuckerPlus$ $(Calculation)$ for cuFastTuckerPlus.
Table \ref{real_dataset_calculation} presents the running time of the aforementioned schemes for cuFastTuckerPlus\_CC and cuFastTuckerPlus on real datasets, while Figure \ref{synthesis_dataset_calculation} illustrates their running time on synthesis datasets. The results indicate that, in the absence of Tensor Cores acceleration, the scheme involving pre-computing $\textbf{C}^{(n)}, n \in\{N\}$ and then reading $\textbf{C}^{(n)}_{\Psi^{(n)},:}, n \in\{N\}$ demonstrates better efficiency. However, when Tensor Cores acceleration is utilized, the real-time calculation of $\textbf{C}^{(n)}_{\Psi^{(n)},:}, n \in\{N\}$ becomes more efficient.
cuFasterTucker reduces computational overhead by introducing additional memory access. Howover, Tensor Cores expedite these calculations, making them faster compared to reading from memory.
This observation highlights that cuFastTuckerPlus outperforms cuFasterTucker and cuFasterTucker\_TC across various aspects, resulting in overall superior performance.

\begin{table}
	\caption{The running time (in seconds) of cuFastTuckerPlus\_CC and cuFastTuckerPlus with various strategies on Netflix and Yahoo!Music datasets.}
	\centering
	\subtable[The process of updating the factor matrices]{
		\centering
		\begin{tabular}{c|ccc}
			\hline \hline
			Algorithm               			& Netflix    & Yahoo!Music   \\
			\hline
			cuFastTuckerPlus\_CC (Calcultion)   & 0.339884   & 0.839495      \\
			cuFastTuckerPlus\_CC (Storage)      & 0.066143   & 0.216421      \\
			cuFastTuckerPlus (Calcultion)       & 0.044434   & 0.146161      \\
			cuFastTuckerPlus (Storage)          & 0.054935   & 0.197515      \\
			\hline \hline
		\end{tabular}
		\label{real_dataset_calculation_factor}
	}
	\qquad
	\subtable[The process of updating the core matrices]{  
		\centering      
		\begin{tabular}{c|ccc}
			\hline \hline
			Algorithm               			& Netflix    & Yahoo!Music   \\
			\hline
			cuFastTuckerPlus\_CC (Calcultion)   & 0.443042   & 1.118276      \\
			cuFastTuckerPlus\_CC (Storage)      & 0.071711   & 0.185181      \\
			cuFastTuckerPlus (Calcultion)       & 0.030067   & 0.072919      \\
			cuFastTuckerPlus (Storage)          & 0.036908   & 0.119635      \\
			\hline \hline
		\end{tabular}
		\label{real_dataset_calculation_core}
	}
	\label{real_dataset_calculation}
\end{table}

\begin{figure}[htbp]
	\centering
	\subfigure[The process of updating the factor matrices]{
		\label{synthesis_dataset_factor}
		\includegraphics[width=1.5in]{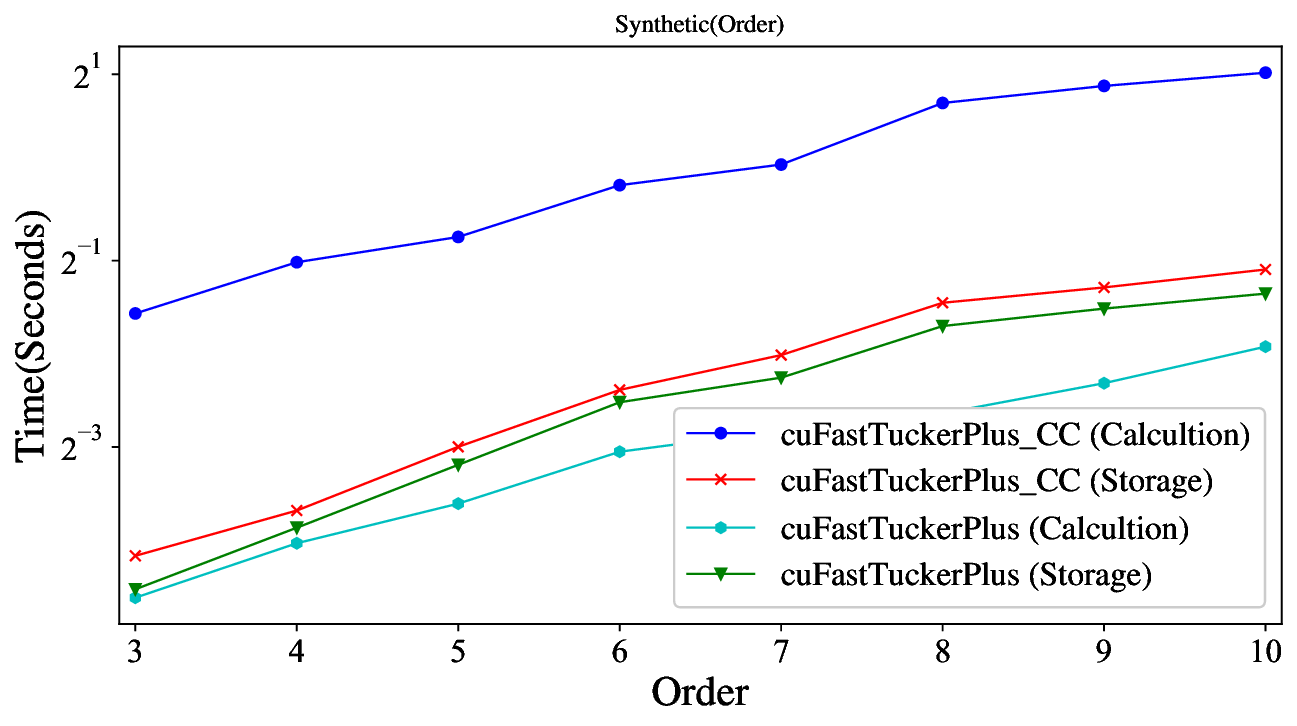}}
	~
	\subfigure[The process of updating the core matrices]{
		\label{synthesis_dataset_core}
		\includegraphics[width=1.5in]{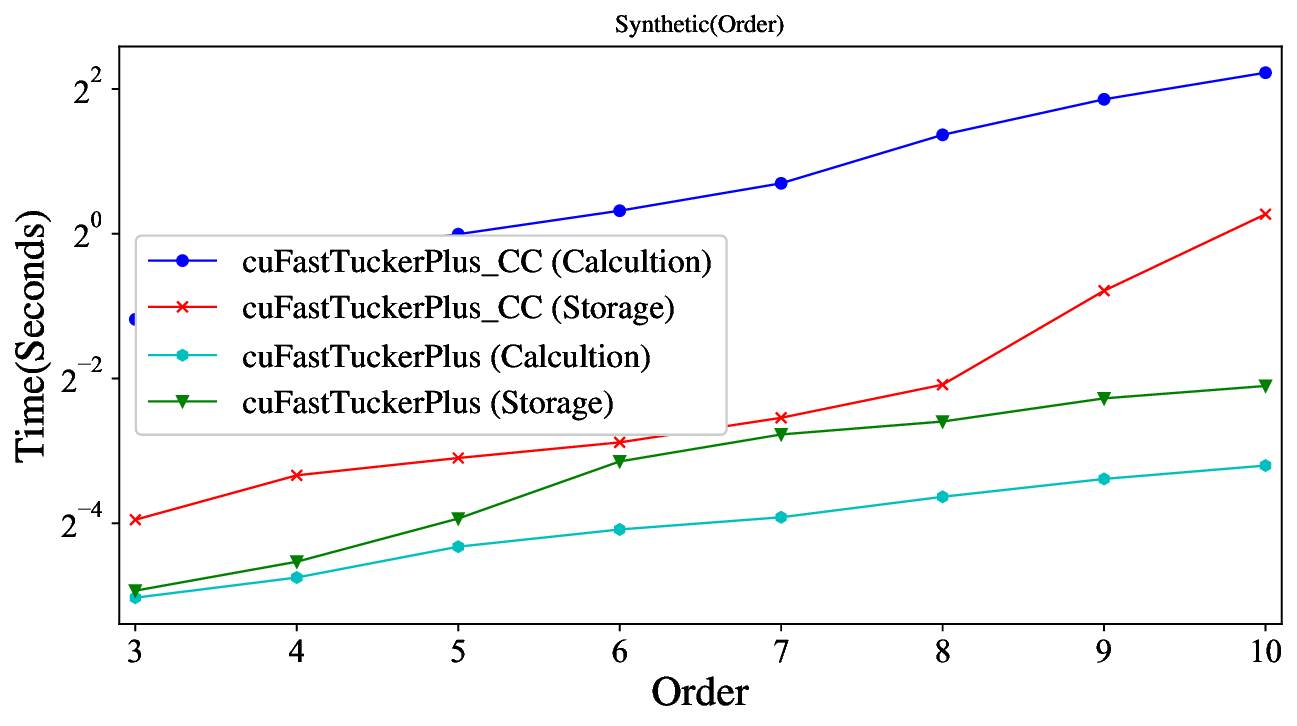}}
	\caption{The running time (in seconds) of cuFastTuckerPlus\_CC and cuFastTuckerPlus in various strategies on the synthesis datasets.}
	\label{synthesis_dataset_calculation}
\end{figure}

\subsection{The Running Time of cuFastTuckerPlus in Various Parameters} \label{Section_The_Running_Time_of_cuFastTuckerPlus_in_Various_Parameters}

The running time of cuFastTuckerPlus is influenced by the values of $R$ and $J_n, n \in \{N\}$. In the same running environment and dataset, a larger value of $R$ or $J_n, n \in \{N\}$ will result in a longer running time for cuFastTuckerPlus. Additionally, when $R$ or $J_n, n \in \{N\}$ is multiplied, the computational complexity of the algorithm also increases approximately.
When $J_n, n \in \{N\}$ is multiplied, it leads to a multiplication of the memory access overhead for $\textbf{A}^{(n)}_{\Psi^{(n)},:}, n \in \{N\}$ and $\textbf{B}^{(n)}, n \in \{N\}$. On the other hand, when $R$ is multiplied, only the memory access overhead for $\textbf{B}^{(n)}, n \in \{N\}$ is multiplied.
Table \ref{The_running_time_in_various_parameters} provides the running time of cuFastTuckerPlus for different parameter configurations: $\big\{R=16, J_n=16, n \in \{N\}\big\}$, $\big\{R=16, J_n=32, n \in \{N\}\big\}$, $\big\{R=32, J_n=16, n \in \{N\}\big\}$ and $\big\{R=32, J_n=32, n \in \{N\}\big\}$ on real datasets.
From the table, it can be observed that when $R$ or $J_n, n \in \{N\}$ is doubled, the running time of cuFastTuckerPlus is less than double the original running time. This can be attributed to two reasons. Firstly, when $R$ increases, the memory access overhead for $\textbf{A}^{(n)}_{\Psi^{(n)},:}, n \in \{N\}$ does not increase. Secondly, proper memory allocation and utilization ensure that the parameter access within a warp does not increase with the increase of $R$ or $J_n, n \in \{N\}$.
These results demonstrate that when the shared memory and registers within a warp are sufficient, a larger value of $R$ or $J_n, n \in \{N\}$ can lead to better cost performance for cuFastTuckerPlus.

\begin{table}
	\caption{The running time (in seconds) of cuFastTuckerPlus in various parameters on Netflix and Yahoo!Music datasets.}
	\label{The_running_time_in_various_parameters}
	\centering
	\begin{tabular}{c|cccc}
		\hline \hline
		&	$R$	&	$J_n$   & 	Netflix    			& Yahoo!Music  	 		\\
		\hline
		\hline
		\multirow{4}*{\makecell[c]{The process of\\ updating the \\factor matrices}}	
		&	16	&	16		&	0.044895			&	0.147708			\\
		&	16	&	32		&	0.087651 (1.95X)	&	0.241832 (1.64X)	\\
		&	32	&	16		&	0.065747 (1.46X)	&	0.193361 (1.31X)	\\
		&	32	&	32		&	0.109624 (2.44X)	&	0.300222 (2.03X)	\\
		\hline
		\multirow{4}*{\makecell[c]{The process of\\ updating the \\core matrices}}
		&	16	&	16		&	0.030488 			&	0.072818  			\\
		&	16	&	32		&	0.049260 (1.62X)	&	0.126714 (1.74X)	\\
		&	32	&	16		&	0.056934 (1.87X)	&	0.139550 (1.92X)	\\
		&	32	&	32		&	0.094193 (3.09X)	&	0.224161 (3.08X)	\\	
		\hline 
		\hline
	\end{tabular}
\end{table}

\section{Conclusion} \label{Section_Conclusion}

We propose the FastTuckerPlus decomposition algorithm, which tackles the entire optimization problem by dividing it into two non-convex optimization subproblems. This approach enables us to leverage the advantages of local search algorithms based on SGD, which exhibit faster convergence compared to global search algorithms used in convex optimization. Additionally, we introduce the cuFastTuckerPlus algorithm, which leverages fine-grained parallelism on GPUs and fully utilizes the capabilities of Tensor Cores.
Our experimental results demonstrate that cuFastTuckerPlus achieves significantly faster convergence compared to existing SOTA algorithms. Specifically, its single iteration time is $3$ to $5$ times faster than the SOTA algorithm, while simultaneously exhibiting smaller memory access overhead. Our algorithm excels at handling HHLST, exhibiting lower memory access and computational overhead compared to existing algorithms. Moreover, it outperforms SOTA algorithms in terms of convergence speed and load balancing.

\section*{Acknowledgment}
This research was supported by the National Key R\&D Program of China (2022ZD0118302).
And the research was partially funded by the Key Program of National Natural Science Foundation of China (Grant Nos. U21A20461, 92055213, 62227808), 
the National Natural Science Foundation of China (Grant No. 61872127).

\ifCLASSOPTIONcaptionsoff
  \newpage
\fi

\bibliographystyle{IEEEtran}
\bibliography{bib}

\begin{thebibliography}{10}
\providecommand{\url}[1]{#1}
\csname url@samestyle\endcsname
\providecommand{\newblock}{\relax}
\providecommand{\bibinfo}[2]{#2}
\providecommand{\BIBentrySTDinterwordspacing}{\spaceskip=0pt\relax}
\providecommand{\BIBentryALTinterwordstretchfactor}{4}
\providecommand{\BIBentryALTinterwordspacing}{\spaceskip=\fontdimen2\font plus
\BIBentryALTinterwordstretchfactor\fontdimen3\font minus
  \fontdimen4\font\relax}
\providecommand{\BIBforeignlanguage}[2]{{%
\expandafter\ifx\csname l@#1\endcsname\relax
\typeout{** WARNING: IEEEtran.bst: No hyphenation pattern has been}%
\typeout{** loaded for the language `#1'. Using the pattern for}%
\typeout{** the default language instead.}%
\else
\language=\csname l@#1\endcsname
\fi
#2}}
\providecommand{\BIBdecl}{\relax}
\BIBdecl

\bibitem{yin2021tensor}
M.~Yin, Y.~Liu, X.~Zhou, and G.~Sun, ``A tensor decomposition based
  collaborative filtering algorithm for time-aware poi recommendation in
  lbsn,'' \emph{Multimedia Tools and Applications}, vol.~80, no.~30, pp.
  36\,215--36\,235, 2021.

\bibitem{wang2021reducing}
J.~Wang, W.~Jiang, K.~Li, and K.~Li, ``Reducing cumulative errors of
  incremental cp decomposition in dynamic online social networks,'' \emph{ACM
  Transactions on Knowledge Discovery from Data (TKDD)}, vol.~15, no.~3, pp.
  1--33, 2021.

\bibitem{wang2019ho}
P.~Wang, L.~T. Yang, G.~Qian, J.~Li, and Z.~Yan, ``Ho-otsvd: A novel tensor
  decomposition and its incremental decomposition for cyber--physical--social
  networks (cpsn),'' \emph{IEEE Transactions on Network Science and
  Engineering}, vol.~7, no.~2, pp. 713--725, 2019.

\bibitem{chen2021deep}
Z.~Chen, Z.~Xu, and D.~Wang, ``Deep transfer tensor decomposition with
  orthogonal constraint for recommender systems,'' in \emph{Proceedings of the
  AAAI Conference on Artificial Intelligence}, vol.~35, no.~5, 2021, pp.
  4010--4018.

\bibitem{shin2022passwordtensor}
Y.~Shin and S.~S. Woo, ``Passwordtensor: Analyzing and explaining password
  strength using tensor decomposition,'' \emph{Computers \& Security}, vol.
  116, p. 102634, 2022.

\bibitem{huang2021tensor}
F.~Huang, X.~Yue, Z.~Xiong, Z.~Yu, S.~Liu, and W.~Zhang, ``Tensor decomposition
  with relational constraints for predicting multiple types of microrna-disease
  associations,'' \emph{Briefings in bioinformatics}, vol.~22, no.~3, p.
  bbaa140, 2021.

\bibitem{zhu2019measuring}
Y.~Zhu, X.~Li, T.~Ristaniemi, and F.~Cong, ``Measuring the task induced
  oscillatory brain activity using tensor decomposition,'' in \emph{ICASSP
  2019-2019 IEEE International Conference on Acoustics, Speech and Signal
  Processing (ICASSP)}.\hskip 1em plus 0.5em minus 0.4em\relax IEEE, 2019, pp.
  8593--8597.

\bibitem{mirzaei2019overlapping}
S.~Mirzaei and H.~Soltanian-Zadeh, ``Overlapping brain community detection
  using bayesian tensor decomposition,'' \emph{Journal of neuroscience
  methods}, vol. 318, pp. 47--55, 2019.

\bibitem{kolda2009tensor}
T.~G. Kolda and B.~W. Bader, ``Tensor decompositions and applications,''
  \emph{SIAM review}, vol.~51, no.~3, pp. 455--500, 2009.

\bibitem{de2000multilinear}
L.~De~Lathauwer, B.~De~Moor, and J.~Vandewalle, ``A multilinear singular value
  decomposition,'' \emph{SIAM journal on Matrix Analysis and Applications},
  vol.~21, no.~4, pp. 1253--1278, 2000.

\bibitem{de2000best}
------, ``On the best rank-1 and rank-(r 1, r 2,..., rn) approximation of
  higher-order tensors,'' \emph{SIAM journal on Matrix Analysis and
  Applications}, vol.~21, no.~4, pp. 1324--1342, 2000.

\bibitem{comon2009tensor}
P.~Comon, X.~Luciani, and A.~L. De~Almeida, ``Tensor decompositions,
  alternating least squares and other tales,'' \emph{Journal of Chemometrics: A
  Journal of the Chemometrics Society}, vol.~23, no. 7-8, pp. 393--405, 2009.

\bibitem{xu2013block}
Y.~Xu and W.~Yin, ``A block coordinate descent method for regularized
  multiconvex optimization with applications to nonnegative tensor
  factorization and completion,'' \emph{SIAM Journal on imaging sciences},
  vol.~6, no.~3, pp. 1758--1789, 2013.

\bibitem{ge2015escaping}
R.~Ge, F.~Huang, C.~Jin, and Y.~Yuan, ``Escaping from saddle points—online
  stochastic gradient for tensor decomposition,'' in \emph{Conference on
  learning theory}.\hskip 1em plus 0.5em minus 0.4em\relax PMLR, 2015, pp.
  797--842.

\bibitem{nisa2019load}
I.~Nisa, J.~Li, A.~Sukumaran-Rajam, R.~Vuduc, and P.~Sadayappan,
  ``Load-balanced sparse mttkrp on gpus,'' in \emph{2019 IEEE International
  Parallel and Distributed Processing Symposium (IPDPS)}.\hskip 1em plus 0.5em
  minus 0.4em\relax IEEE, 2019, pp. 123--133.

\bibitem{oh2018scalable}
S.~Oh, N.~Park, S.~Lee, and U.~Kang, ``Scalable tucker factorization for sparse
  tensors-algorithms and discoveries,'' in \emph{2018 IEEE 34th International
  Conference on Data Engineering (ICDE)}.\hskip 1em plus 0.5em minus
  0.4em\relax IEEE, 2018, pp. 1120--1131.

\bibitem{park2021vest}
M.~Park, J.-G. Jang, and L.~Sael, ``Vest: Very sparse tucker factorization of
  large-scale tensors,'' in \emph{2021 IEEE International Conference on Big
  Data and Smart Computing (BigComp)}.\hskip 1em plus 0.5em minus 0.4em\relax
  IEEE, 2021, pp. 172--179.

\bibitem{li2020sgd}
H.~Li, Z.~Li, K.~Li, J.~S. Rellermeyer, L.~Chen, and K.~Li, ``Sgd\_\_tucker: A
  novel stochastic optimization strategy for parallel sparse tucker
  decomposition,'' \emph{IEEE Transactions on Parallel and Distributed
  Systems}, vol.~32, no.~7, pp. 1828--1841, 2020.

\bibitem{oh2019high}
S.~Oh, N.~Park, J.-G. Jang, L.~Sael, and U.~Kang, ``High-performance tucker
  factorization on heterogeneous platforms,'' \emph{IEEE Transactions on
  Parallel and Distributed Systems}, vol.~30, no.~10, pp. 2237--2248, 2019.

\bibitem{li2022cu_fasttucker}
Z.~Li, ``cu\_fasttucker: A faster and stabler stochastic optimization for
  parallel sparse tucker decomposition on multi-gpus,'' \emph{arXiv preprint
  arXiv:2204.07104}, 2022.

\bibitem{li2022cufastertucker}
------, ``cufastertucker: A stochastic optimization strategy for parallel
  sparse fasttucker decomposition on gpu platform,'' \emph{arXiv preprint
  arXiv:2210.06014}, 2022.

\bibitem{ma2021local}
T.~Ma, ``Why do local methods solve nonconvex problems?'' \emph{Beyond the
  Worst-Case Analysis of Algorithms}, p. 465, 2021.

\bibitem{dauphin2014identifying}
Y.~N. Dauphin, R.~Pascanu, C.~Gulcehre, K.~Cho, S.~Ganguli, and Y.~Bengio,
  ``Identifying and attacking the saddle point problem in high-dimensional
  non-convex optimization,'' \emph{Advances in neural information processing
  systems}, vol.~27, 2014.

\bibitem{choromanska2015loss}
A.~Choromanska, M.~Henaff, M.~Mathieu, G.~B. Arous, and Y.~LeCun, ``The loss
  surfaces of multilayer networks,'' in \emph{Artificial intelligence and
  statistics}.\hskip 1em plus 0.5em minus 0.4em\relax PMLR, 2015, pp. 192--204.

\bibitem{park2017non}
D.~Park, A.~Kyrillidis, C.~Carmanis, and S.~Sanghavi, ``Non-square matrix
  sensing without spurious local minima via the burer-monteiro approach,'' in
  \emph{Artificial Intelligence and Statistics}.\hskip 1em plus 0.5em minus
  0.4em\relax PMLR, 2017, pp. 65--74.

\bibitem{bhojanapalli2016global}
S.~Bhojanapalli, B.~Neyshabur, and N.~Srebro, ``Global optimality of local
  search for low rank matrix recovery,'' \emph{Advances in Neural Information
  Processing Systems}, vol.~29, 2016.

\bibitem{ge2016matrix}
R.~Ge, J.~D. Lee, and T.~Ma, ``Matrix completion has no spurious local
  minimum,'' \emph{Advances in neural information processing systems}, vol.~29,
  2016.

\bibitem{ge2017no}
R.~Ge, C.~Jin, and Y.~Zheng, ``No spurious local minima in nonconvex low rank
  problems: A unified geometric analysis,'' in \emph{International Conference
  on Machine Learning}.\hskip 1em plus 0.5em minus 0.4em\relax PMLR, 2017, pp.
  1233--1242.

\bibitem{frandsen2022optimization}
A.~Frandsen and R.~Ge, ``Optimization landscape of tucker decomposition,''
  \emph{Mathematical Programming}, vol. 193, no.~2, pp. 687--712, 2022.

\bibitem{carmon2018accelerated}
Y.~Carmon, J.~C. Duchi, O.~Hinder, and A.~Sidford, ``Accelerated methods for
  nonconvex optimization,'' \emph{SIAM Journal on Optimization}, vol.~28,
  no.~2, pp. 1751--1772, 2018.

\bibitem{agarwal2016finding}
N.~Agarwal, Z.~Allen-Zhu, B.~Bullins, E.~Hazan, and T.~Ma, ``Finding
  approximate local minima for nonconvex optimization in linear time,''
  \emph{arXiv preprint arXiv:1611.01146}, 2016.

\bibitem{jin2017escape}
C.~Jin, R.~Ge, P.~Netrapalli, S.~M. Kakade, and M.~I. Jordan, ``How to escape
  saddle points efficiently,'' in \emph{International conference on machine
  learning}.\hskip 1em plus 0.5em minus 0.4em\relax PMLR, 2017, pp. 1724--1732.

\bibitem{feng2021apnn}
B.~Feng, Y.~Wang, T.~Geng, A.~Li, and Y.~Ding, ``Apnn-tc: Accelerating
  arbitrary precision neural networks on ampere gpu tensor cores,'' in
  \emph{Proceedings of the international conference for high performance
  computing, networking, storage and analysis}, 2021, pp. 1--13.

\bibitem{finkelstein2021quantum}
J.~Finkelstein, J.~S. Smith, S.~M. Mniszewski, K.~Barros, C.~F. Negre, E.~H.
  Rubensson, and A.~M. Niklasson, ``Quantum-based molecular dynamics
  simulations using tensor cores,'' \emph{Journal of Chemical Theory and
  Computation}, vol.~17, no.~10, pp. 6180--6192, 2021.

\bibitem{huang2022high}
H.~Huang, X.-Y. Liu, W.~Tong, T.~Zhang, A.~Walid, and X.~Wang, ``High
  performance hierarchical tucker tensor learning using gpu tensor cores,''
  \emph{IEEE Transactions on Computers}, 2022.

\bibitem{li2020accelerating}
A.~Li and S.~Su, ``Accelerating binarized neural networks via bit-tensor-cores
  in turing gpus,'' \emph{IEEE Transactions on Parallel and Distributed
  Systems}, vol.~32, no.~7, pp. 1878--1891, 2020.

\bibitem{zhu2019sparse}
M.~Zhu, T.~Zhang, Z.~Gu, and Y.~Xie, ``Sparse tensor core: Algorithm and
  hardware co-design for vector-wise sparse neural networks on modern gpus,''
  in \emph{Proceedings of the 52nd Annual IEEE/ACM International Symposium on
  Microarchitecture}, 2019, pp. 359--371.

\bibitem{zachariadis2020accelerating}
O.~Zachariadis, N.~Satpute, J.~G{\'o}mez-Luna, and J.~Olivares, ``Accelerating
  sparse matrix--matrix multiplication with gpu tensor cores,'' \emph{Computers
  \& Electrical Engineering}, vol.~88, p. 106848, 2020.

\bibitem{mukunoki2020dgemm}
D.~Mukunoki, K.~Ozaki, T.~Ogita, and T.~Imamura, ``Dgemm using tensor cores,
  and its accurate and reproducible versions,'' in \emph{High Performance
  Computing: 35th International Conference, ISC High Performance 2020,
  Frankfurt/Main, Germany, June 22--25, 2020, Proceedings}.\hskip 1em plus
  0.5em minus 0.4em\relax Springer, 2020, pp. 230--248.

\bibitem{firoz2020feasibility}
J.~S. Firoz, A.~Li, J.~Li, and K.~Barker, ``On the feasibility of using
  reduced-precision tensor core operations for graph analytics,'' in \emph{2020
  IEEE High Performance Extreme Computing Conference (HPEC)}.\hskip 1em plus
  0.5em minus 0.4em\relax IEEE, 2020, pp. 1--7.

\end{thebibliography}

\end{document}